\documentclass[epj]{msvjour}
\usepackage{latexsym}
\usepackage{graphics}
\usepackage{epsfig}
\usepackage{amsfonts}
\begin{document}
\newcommand*{\pd}{\partial}
\newcommand*{\pdm}{\pd_{\mu}}
\newcommand*{\pdn}{\pd_{\nu}}
\newcommand*{\pdi}{\pd_{i}}
\newcommand*{\pda}[1]{\pd_#1}
\newcommand*{\be}{\begin{equation}}
\newcommand*{\ee}{\end{equation}}
\newcommand*{\bea}{\begin{eqnarray}}
\newcommand*{\eea}{\end{eqnarray}}
\newcommand*{\pref}[1]{(\ref{#1})}
\title{High-Temperature Limit of Landau-Gauge Yang-Mills Theory}
\author{Axel Maas\inst{1} \and Jochen Wambach\inst{1,2} \and Burghard Gr\"uter \inst{3} \and Reinhard Alkofer \inst{3}}     
\institute{Institute for Nuclear Physics, Darmstadt University of Technology, Schlo{\ss}gartenstra{\ss}e 
9, D-64289 Darmstadt, Germany \and Gesellschaft f\"ur Schwerionenforschung mbH, Planckstr. 1, 
D-64291 Darmstadt, Germany 
\and Institute of Theoretical Physics, T\"ubingen University, Auf der Morgenstelle 14, 
D-72076 T\"ubingen, Germany}
\date{\today}
\abstract{The infrared properties of the high-temperature limit of Landau-gauge 
Yang-Mills theory are investigated. In a first step the high-temperature
limit of the Dyson-Schwinger equations is taken. The resulting
equations are identical to the Dyson-Schwinger equations of the dimensionally
reduced theory, a three-dimensional Yang-Mills theory coupled to an effective
adjoint Higgs field. These equations are solved analytically in the infrared
and ultraviolet, and numerically for all Euclidean momenta. We find infrared
enhancement for the Faddeev-Popov ghosts, infrared suppression for transverse
gluons and a mass for the Higgs. These results imply long-range interactions
and over-screening in the chromomagnetic sector of high temperature Yang-Mills
theory while in the chromoelectric sector only screening is observed.
} 
\maketitle
\section{Introduction}\label{intro}

The infrared structure of Quantum Chromo Dynamics (QCD) governs the dynamics of
hadrons as well as the thermodymanic properties of hot and dense hadronic matter
and is thus of great current interest. Although our understanding is far from
being satisfactory much progress has been made recently using different
genuinely non-per\-turbative techniques. Such methods include lattice Monte-Carlo
calculations, Dyson-Schwinger equations (DSEs),\linebreak renormalization group methods,
stochastic quantization, the use of topological arguments, and others.

For QCD, being a gauge theory, it is expected that the description of strong
interaction phenomena such as confinement contains gauge dependent aspects.
Therefore investigations in a particular gauge will very likely not
lead to a full understanding of non-perturbative dynamics but may nevertheless
provide important information. It turns out that for the studies to be
performed here the Landau gauge is advantageous~\cite{Alkofer:2003jr} due
to its non-renormalization of the ghost-gluon vertex~\cite{Taylor:ff}.

One of the most intriguing phenomena measured with extremely high precision is
confinement, {\it i.e.\/} the absence of colored objects from the physical
spectrum. This is a genuinely non-perturbative effect related to the infrared
behavior of QCD. Monte-Carlo lattice calculations provide clear evidence that
the confining properties of QCD change above a critical temperature, see {\it
e.g.\/}~\cite{Karsch:2003jg}. This has implications for relativistic heavy-ion
collisions and the early stages of the universe. Since it is likely that
confinement is generated in the Yang-Mills sector of QCD, the behavior of a
pure Yang-Mills theory may already reveal the qualitative mechanisms of the
deconfinement transition. It is therefore especially interesting to study the
properties of Yang-Mills theory above the critical temperature as well as its
temperature dependence.

Here we will present the results of such a study. For technical reasons, this
is currently done in the infinite-temperature limit, but investigations with
finite-temp\-er\-at\-ure corrections are ongoing. (The results of an earlier
ex\-ploratory study are presented in~\cite{Maas:2002if}.)

Complementary to this approach also a study at temperatures below the phase
transition has been performed. The corresponding results will be presented 
elsewhere
\cite{torus}. Both, these and the investigations presented here utilize DSEs
\cite{Dyson:1949ha}, the equations of motion of the Yang-Mills theory. They
therefore extend successful calculations in the vacuum of the Yang-Mills theory
\cite{vonSmekal:1997is} and full QCD~\cite{Fischer:2003rp} to finite
temperature.

This paper is organized as follows: To make the presentation self-contained we
briefly review some aspects of confinement in Yang-Mills theories in section
\ref{Confinement}. In section \ref{DSE} we derive the DSEs to be solved and
discuss the truncations made. In section \ref{Analytic} we present analytic
solutions for different momentum regimes while section \ref{Numerical} exhibits
the full numerical solutions as well as their comparison with lattice
calculations. In section \ref{Derived} we discuss the thermodynamic potential
and screening masses. Interpretations of the results with respect to different
aspects are given in section \ref{Interpretation}. In section \ref{Conclusions}
we conclude and also give an outlook concerning ongoing activities.
Technical details are deferred to five appendices.

\section{Confinement}\label{Confinement}

A major focus of the present work is the fate of confinement at temperatures
far above the phase transition. It is therefore necessary to be able to extract
information about confinement. To establish criteria for confinement
we study the infrared behavior of the
pertinent 2-point functions.\footnote{ In a $SU(N)$ Yang-Mills theory 1-point 
functions are expected to vanish even at
non-vanishing temperatures due to the antisymmetry of the structure constants. 
No sign of a symmetric color structure of vertex functions, although in
principle possible for $SU(N\ge 3)$ gauge theories, has yet been found~\cite{Boucaud:1998xi}. In the
following we therefore assume it not to be present. The dimensionally reduced
theory, as the high-temperature limit of a four-dimensional Yang-Mills theory, 
then also only contains color antisymmetric vertex functions.}  Since 
calculations in Euclidian space-time are significantly simpler than in Minkowski 
space-time  we use in the following three criteria which are applicable in the 
Euclidian case.\footnote{Employing Euclidian space-time also facilitates comparisons 
with lattice results. The transformation back to Min\-kowski space-time is already
non-trivial in the vacuum~\cite{Alkofer:2003jj} and will be addressed in
section \ref{analytic}. For the case of equilibrium calculations this point is
of minor concern since the statistical system is anyway described in 
Euclidian space.}

The first criterion is an empirical one. It allows to infer that a particle
is confined irrespective of the dynamical origin of confinement. It is based on
the fact that no K\"all{\'e}n-Lehmann representation of a particle exists if
the corresponding propagator does not have a positive semi-definite spectral
norm, {\it i.e.\/} for this ``particle'' the Osterwalder-Schrader axiom of
reflection positivity~\cite{Osterwalder:dx} is violated. It is then not part of
the physical spectrum and thus confined, {\it c.f.\/} ref.\ \cite{Oehme:bj}. To
test this criterion we will investigate whether the Schwinger function $\Delta(t)$ 
is positive semi-definite, see section \ref{Masses} below. In case the
corresponding propagator $D$ vanishes at zero momentum,
\be 
\lim_{q^2 \to 0}D(q^2)=0 ,\label{Oehme} 
\ee 
the respective Schwinger function cannot be positive, and thus positivity 
is violated. It is important to note that, since positivity
violation is only a sufficient condition for confinement, particles may still
be confined by other mea\-ns.

The two remaining criteria describe confinement mechanisms. The Kugo-Ojima
scenario~\cite{Kugo:gm} puts forward the idea that all colored objects form
BRST-quartets and thus do not belong to the physical state space. This scenario
is based on a derivation requiring three preconditions which have (yet) to
be proven. The first precondition is an unbroken BRST charge also for large
and thus non-per\-turbative field configurations. The second is the failure of
the cluster decomposition theorem. This requires that no mass gap exists in the
complete state space (although a mass gap has to exist in the physical
subspace). The third requirement is an unbroken global color charge. Assuming
the validity of these conjectures, the Kugo-Ojima confinement criterion is
fulfilled in the Landau gauge if~\cite{Kugo:1995km}
\be 
\lim_{q^2 \to 0} q^2 D_G(q^2)\to\infty,
\label{Kugo} 
\ee
where $D_G$ is the propagator of the Faddeev-Popov ghosts. This scenario 
necessarily also implies the condition (\ref{Oehme}) for the gluon propagator.

In the Zwanziger-Gribov scenario~\cite{Zwanziger:2003cf}, entropy arguments are 
employed to show the dominance of field configurations close or on the Gribov
horizon in field configuration space. It has two consequences that can be
investigated. This scenario predicts a behavior of the gluon propagator as in
condition (\ref{Oehme}) and of the ghost propagator as in condition
(\ref{Kugo}). In addition, it is argued that the gauge fixing term  dominates
the action. Therefore the ghost-loop-only truncation, introduced below, becomes
exact in the infrared.  Our results, as the ones obtained in the vacuum~\cite{vonSmekal:1997is,Fischer:2003rp}, support this picture in so far as 
(\ref{Oehme}) and (\ref{Kugo}) are satisfied and that in the infrared limit the 
behavior of the gluon propagator is driven by the ghost loop. 

Also intuitively it is clear that a strongly divergent ghost propagator at zero
momentum can mediate confinement. Such an infrared divergence relates to long-ranged 
spatial correlations. These are stronger than the ones induced by a Coulomb force 
since the divergence in momentum space is stronger than that of a massless particle.

\section{Dyson-Schwinger Equations}\label{DSE}

\subsection{Tensor structure of the gluon equation}

The DSEs~\cite{Dyson:1949ha,Rivers:hi,Alkofer:2000wg,Roberts:2000aa} form an infinite 
tower of coupled non-linear integral equations for the Green's function of a given
theory. Thus it is necessary to truncate this system. The motivation for the specific
truncation scheme used here will be given below. In the following we aim at a closed 
set of equations for the pertinent two-point functions. In Landau gauge and 
for the vacuum in 3+1 dimensions these are the ghost propagator
\be
D_G(q^2)=\frac{-G(q^2)}{q^2}\label{ghostDressing}
\ee
and the gluon propagator
\be
D_{\mu\nu}(q)=P_{\mu\nu}(q)\frac{Z(q^2)}{q^2}\label{gluonDressing}
\ee
where $P_{\mu\nu}(q)=\delta_{\mu\nu}-q_\mu q_\nu /q^2$ is the transverse
projector. Eqs.\ (\ref{ghostDressing}) and (\ref{gluonDressing}) define the 
dimensionless dressing functions $G(q^2)$ and $Z(q^2)$, respectively.

\begin{figure}
\epsfig{file=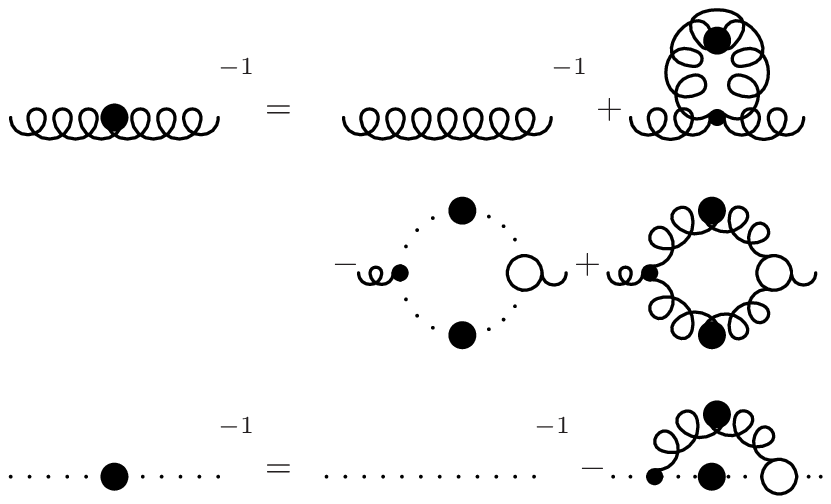,width=0.95\linewidth}
\caption{Graphical representation of the truncated system of vacuum DSEs. 
Wiggly lines represent
gluons and dotted lines ghosts. Lines with a full dot are full propagators. Small 
dots denote bare vertices and open circles represent full vertices.}
\label{4dtrunc}
\end{figure}

We will allow, but not require, ghost-loop dominance in the infrared. In the 
ultraviolet the one-loop terms are dominant as they are the leading terms in 
perturbation theory. Thus we follow here ref.\ \cite{vonSmekal:1997is} and neglect
one-parti\-cle-irreducible two-loop diagrams.  The derivation of these equations
at zero temperature can be found {\it e.g.} in~\cite{Alkofer:2000wg,Roberts:2000aa}. 
The tadpole is not left as a free part of the equations but its non-perturbative 
behavior is fixed by the requirement that no divergences beyond those allowed by the
Slavnov-Taylor identities should occur in the gluon equation. This reflects the
behavior of the tadpole in perturbative calculations. This truncation scheme is
depicted in Fig.\ \ref{4dtrunc}. 

Finite temperature is introduced in the DSEs using the Matsubara formalism~\cite{Das:gg}. 
This entails two independent tensor structures for the gluon,
one of them being three-dimensional (3d) transverse and the other one 
3d longitudinal~\cite{Kapusta:tk}
\be
D_{\mu\nu}(q)=P_{T\mu\nu}(q)\frac{Z(q_0^2,\vec q^2)}{q^2}+
P_{L\mu\nu}(q)\frac{H(q_0^2,\vec q^2)}{q^2} \; .
\ee
Therefore there are two independent dressing functions, $Z$ and 
$H$.\footnote{These are identically to $Z_T$ and $Z_L$ in~\cite{Maas:2002if},
respectively. The notation has been changed to better visualize their 
correspondence to the objects in the 3d theory. In~\cite{torus}
they are called $Z_M$ and $Z_0$ to emphasize their physical meaning.} 
The variable $q_0$ denotes the Matsubara frequency and $\vec q$ the spatial momentum.
Note that Lorentz invariance, although not manifestly visible any more, is not
lost in this formalism~\cite{Weldon:aq}. The ghost propagator, as a scalar, does 
not acquire a second independent dressing  function. It depends nevertheless also 
on $q_0$ and $\vec q$ separately. Therefore three independent functions of two 
variables each have to be determined.

\begin{figure}
\epsfig{file=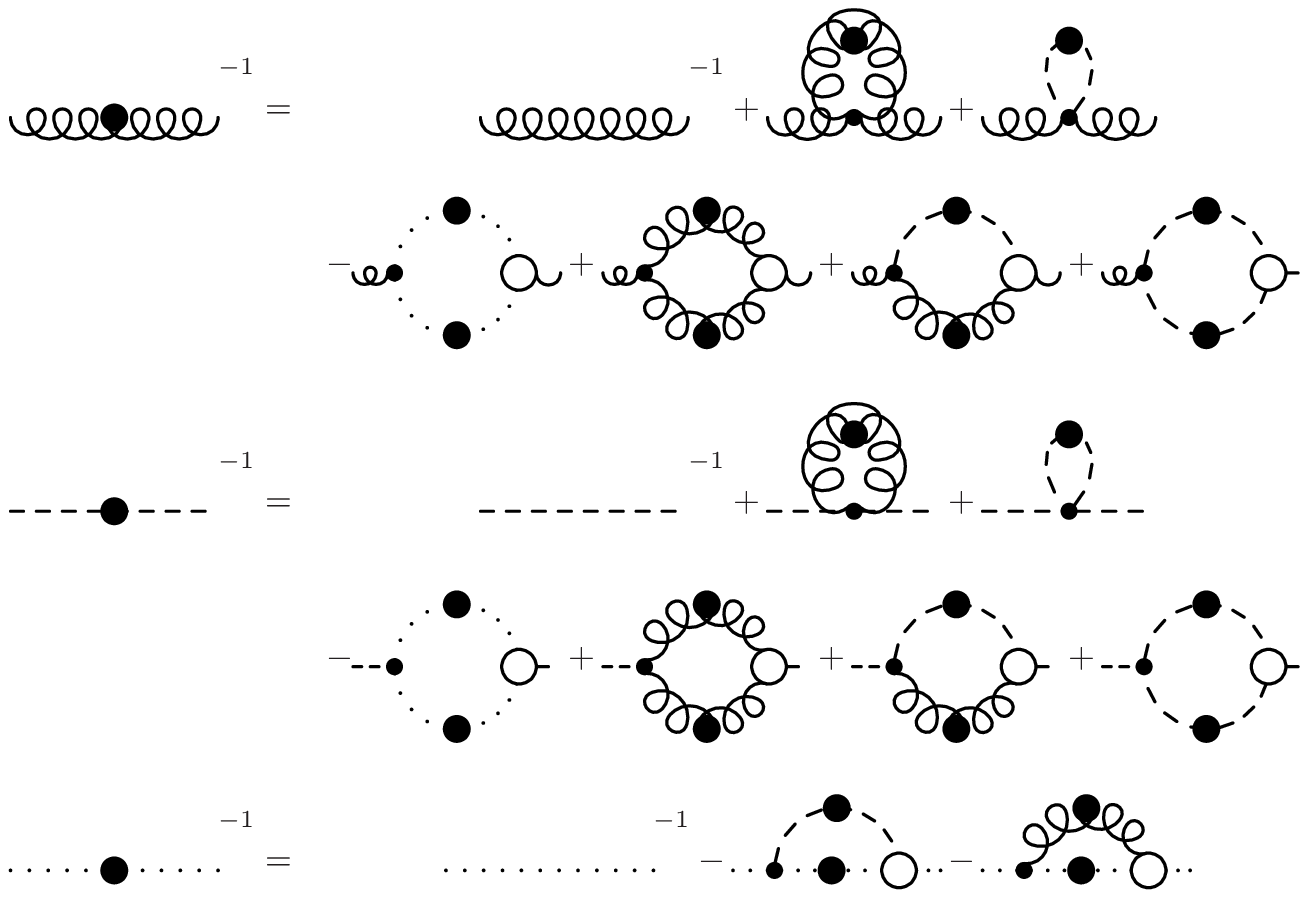,width=0.95\linewidth}
\caption{Graphical representation of the truncated system of DSEs at 
non-vanishing temperature. Wiggly lines represent 3d transverse gluons, 
dashed lines 3d longitudinal gluons and dotted lines ghosts. 
Lines with a full dot are full propagators. Small dots denote bare vertices and 
open circles represent the full vertices which have to be constructed in the 
truncation discussed in the text.}
\label{ftsys}
\end{figure}

To obtain from the matrix equation for the gluon propagator two scalar
equations for the dressing functions, it is contracted with two tensors~\cite{Maas:2002if}
\bea
P_{L\mu \nu }^{\xi }&=&\xi P_{L\mu \nu }+
(1-\xi )(1+\frac{q_{0}^{2}}{\vec q^{2}})\delta _{\mu 0}\delta _{0\nu }\nonumber\\
P_{T\mu \nu }^{\zeta }&=&\zeta P_{T\mu \nu }+
(1-\zeta )(\delta _{\mu \nu }-(1+\frac{q_{0}^{2}}{\vec q^{2}})
\delta _{\mu 0}\delta _{0\nu }).\label{project}
\eea
where $\xi$ and $\zeta$ are two parameters. They are connected to the
truncation scheme and will also be discussed in subsection \ref{Truncation}.
$P_L$ and $P_T$ are defined as~\cite{Kapusta:tk}
\bea
P_{T\mu \nu }&=&\delta _{\mu \nu }-
\frac{q_{\mu }q_{\nu }}{\vec q^{2}}+\nonumber\\
&&+\delta _{\mu 0}\frac{q_{0}q_{\nu }}{\vec q^{2}}
+\delta _{0\nu }\frac{q_{\mu }q_{0}}{\vec q^{2}}
-\delta _{\mu 0}\delta _{0\nu }(1+\frac{q_{0}^{2}}{\vec q^{2}})\\
P_{L\mu \nu }&=&P_{\mu \nu }-P_{T\mu \nu }.
\eea
Graphically these equations are represented in Fig.\ \ref{ftsys}. 
The solutions for small temperatures below the phase 
transition are discussed in~\cite{torus}.

\subsection{Dimensional Reduction}\label{Reduction}

In the next step we take the infinite temperature limit of the truncated DSEs.
To this end we note that, neglecting any contributions from Matsubara
frequencies different from zero, this yields an effective three-dimensional
Yang-Mills theory with an additional adjoint Higgs. This Higgs field is the
remnant of the $A_0$ field in the four-dimensional theory. Thus the number
of degrees of freedom is conserved.

The structure of the projectors (\ref{project}) simplifies in the case of
vanishing Matsubara frequency, {\it i.e.} $q_0=0$. $P_{L\mu\nu}^\xi$ becomes $\delta_{\mu 0}\delta_{0\nu}$
independent of $\xi$. It thus projects onto the time-time component of the
propagator. Therefore, the 3d longitudinal part relates to the Higgs and this
sector corresponds to the chromoelectric sector of the original
theory. The non-vanishing elements of $P_{T\mu\nu}^\zeta$ become 
\be
P_{ij}^\zeta=\delta_{ij}-\zeta\frac{p_i p_j}{p^2} \; .\label{bpproj}
\ee
It projects onto the three-dimensional subspace. Therefore the
gluons in the 3d reduced theory correspond to the chromomagnetic
sector of the four-dimensional theory.

This dimensionally reduction amounts to integrating out the hard modes in 
tree-level approximation. Contributions from higher Matsubara frequencies can 
potentially generate additional terms in the Lagrangian of the dimensionally 
reduced theory~\cite{Kajantie:1995dw}. Lattice calculations and matching in the 
perturbative regime show that this is indeed the case~\cite{Kajantie:1995dw,Cucchieri:2001tw}. Especially, a tree-level mass for the Higgs is
generated. This also modifies the DSEs: the additional terms occurring stem from 
the neglected Matsubara frequencies and once included they can be derived from 
first principles as follows. The Lagrangian which governs the 3d reduced theory 
is \cite{Alkofer:2000wg,Kajantie:1995dw}
\bea
{\cal L}&=&\frac{1}{4}F_{\mu\nu}^aF_{\mu\nu}^a
+\bar c^a \pdm D_\mu^{ab} c^b\nonumber\\
&+&\frac{1}{2}(D_\mu^{ab}\phi^b D_\mu^{ac}\phi^c
+m_h^2\phi^a\phi^a)+\frac{h}{4}\phi^a\phi^a\phi^b\phi^b\label{lagrange}
\eea
with the field strength tensor $F_{\mu\nu}^a$ and the covariant derivative
$D_\mu^{ab}$ defined as
\bea
F^a_{\mu\nu}&=\pdm A_\nu^a-\pdn A_\mu^a-g_3f^{abc}A_\mu^bA_\nu^c\\
D_\mu^{ab}&=\delta^{ab}\pdm+g_3f^{abc}A_\mu^c \; .
\eea
$A_\mu^a$ is the gauge field, $c^a$ the ghost, $\bar c^a$ the
anti-ghost\footnote{Note that the hermiticity assignment for the ghost is not
the usual one but is still correct for Landau gauge~\cite{Alkofer:2003jr,Alkofer:2000wg}.}, $\phi^a$ the Higgs field, $g_3$ the
dimensionful gauge coupling, $m_h$ is the Higgs mass, $h$ is the Higgs
self-coupling and $f^{abc}$ the structure constants of the gauge group. A
three-Higgs coupling is not present due to the antisymmetry of the structure
constants and the remaining global color symmetry at tree-level.

If appropriate, multiples of $g_3^2$ are chosen as the fundamental scale and all
results in the employed truncation scheme will be independent of $N_c$ for
gauge groups SU($N_c$) in the following sense: 
In the chosen truncation scheme only the combination $C_Ag_3^2$ appears as
reference to the gauge group, $C_A=N_c$ being the second Casimir invariant 
of the adjoint representation of the group. Hence a change in $N_c$ can
always be absorbed by a change in $g_3^2$ as long as the ratio $g_3^2/m_h$ is 
kept fixed.

The Higgs self-coupling $h$ can be uniquely determined in the 3d reduced
theory. Since the Higgs field is a component of the four-dimensional gluon field, 
no linear divergence of the Higgs self-energy can occur due to the
Slavnov-Taylor identities of the four-dimensional theory. Introducing
temperature, no novel divergencies arise~\cite{Das:gg}. Implementation of this
requirement fixes the Higgs self-coupling $h$ in leading-order perturbation 
theory (see appendix \ref{appUV}).   

All other constants occurring in the Lagrangian (\ref{lagrange}) are effective
constants which arise by integrating out the heavy modes. For any large,
although finite, temperature $T$ all remaining dimensionful quantities will
scale as the temperature $T$. After rescaling, the precise dependence of $g_3$ 
on $T$ is irrelevant for the solution of the DSEs and the issue will be postponed 
to subsection \ref{td}.

The 3d reduced theory is not only superrenormalizable but finite, and all
renormalization constants will be set to one in the following.  

\begin{figure}
\epsfig{file=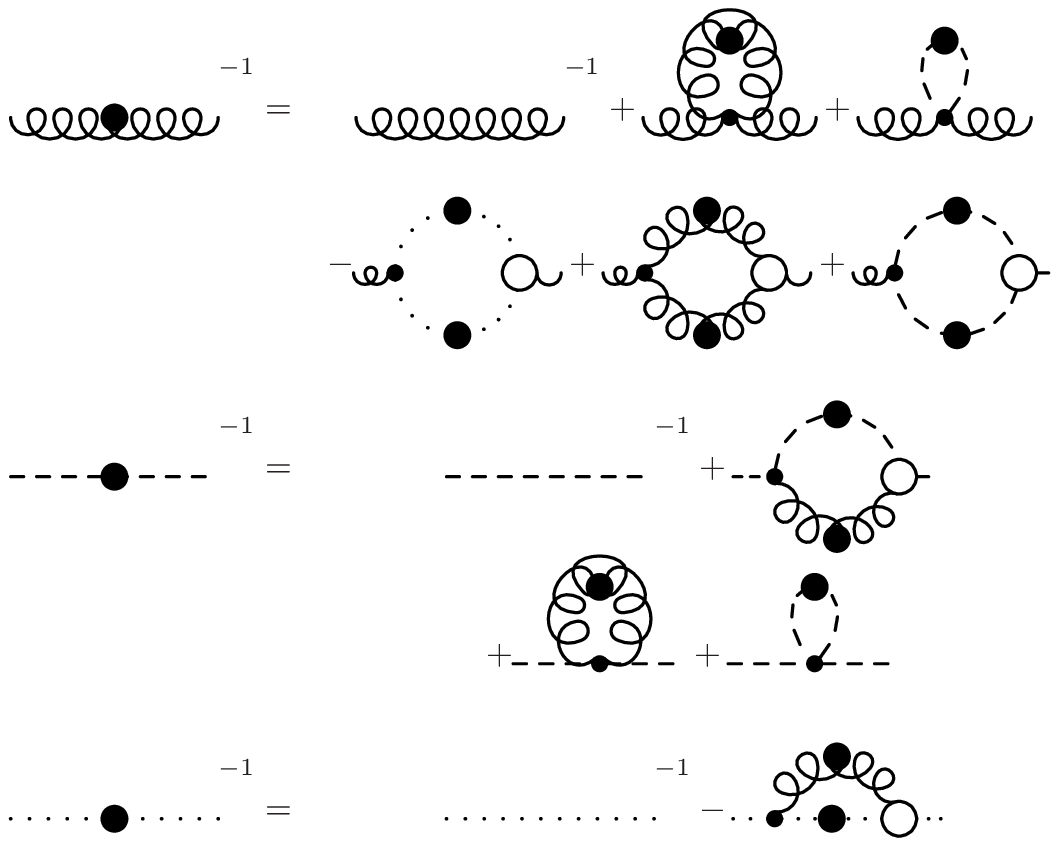,width=0.95\linewidth}
\caption{Graphical representation of the truncated system of DSEs 
in the infinite temperature limit. Wiggly lines represent gluons,
dashed lines Higgs and dotted lines ghosts. Lines with a full dot are full 
propagators. Small dots denote bare vertices and open circles represent full vertices.}
\label{threedsys}
\end{figure}

The DSEs for the Yang-Mills sector are already known, see {\it e.g.}~\cite{Alkofer:2000wg}. 
They are rederived in appendix \ref{appdse} for completeness. Since no tree level 
coupling between the Higgs and the ghost is present, the ghost equation will not 
be modified as compared to pure Yang-Mills theory. The gho\-st and the gluon DSEs 
are given by
\bea
(D_G^{-1})^{ab}(p)&=&-\delta^{ab}p^2\nonumber\\
&+&\int\frac{d^dq}{(2\pi)^d}
\Gamma_\mu^{tl;c\bar cA;dae}(-q,p,q-p)D^{ef}_{\mu\nu}(p-q)
\nonumber\\
&&\qquad \times D^{dg}_G(q) \Gamma^{c\bar cA;bgf}_\nu(-p,q,p-q) \; ,
\label{ghost}
\\ \nonumber\\
(D_H^{-1})^{ab}(p)&=&\delta^{ab}(p^2+m_h^2)+T^{HG;ab}+T^{HH;ab}\nonumber\\
&+&\int\frac{d^dq}{(2\pi)^d}\Gamma^{tl;A\phi^2;eac}_\nu(-p-q,p,q)
D_{\nu\mu}^{cg}(p+q)\nonumber\\
&&\qquad \times D^{fc}(q)\Gamma_\mu^{gbf}(p+q,-p,-q) \; , \label{higgs}
\\ \nonumber\\
(D^{-1})^{ab}_{\mu\nu}(p)&=&\delta^{ab}(\delta_{\mu\nu}p^2-p_\mu p_\nu)+
T_{\mu\nu}^{GG;ab}+T^{GH;ab}_{\mu\nu}\nonumber\\
&-&\int\frac{d^dq}{(2\pi)^d}\Gamma_\mu^{tl;c\bar cA;dca}(-p-q,q,p) D_G^{cf}(q) 
\nonumber\\
&&\qquad \times D_G^{de}(p+q) \Gamma_\nu^{c\bar c A;feb}(-q,p+q,-p)\nonumber\\
&+&\frac{1}{2}\int\frac{d^dq}{(2\pi)^d}\Gamma^{tl;A^3;acd}_{\mu\sigma\chi}(p,q-p,-q)
D^{cf}_{\sigma\omega}(q)\nonumber\\
&&\qquad \times D^{de}_{\chi\lambda}(p-q) \Gamma^{A^3;bfe}_{\nu\omega\lambda}(-p,q,p-q)\nonumber\\
&+&\frac{1}{2}\int\frac{d^dq}{(2\pi)^d}\Gamma_\mu^{tl;A\phi^2;acd}(p,q-p,-q)D^{de}(q)\nonumber\\
&&\qquad \times D^{cf}(p-q)\Gamma^{A\phi^2;bef}_\nu(-p,q,p-q) \; ,
\nonumber\\ \label{gluon}
\eea
where $T^i$ are the tadpole contributions. The first label gives the equation
where the tadpole contributes ($G$ for gluon and $H$ for Higgs) and the second
the type of tadpole appearing. The index $tl$ denotes the tree-level quantities 
which are explicitely given in eqs.\  (\ref{tlcca}) - (\ref{tl4h}) in appendix
\ref{appdse}. This set of 3d truncated equations is diagrammatically displayed 
in Fig.~\ref{threedsys}. Note that all momenta in these equations are
three-momenta.

It is important to note that the ghost-gluon vertex function 
$\Gamma^{c\bar cA;abc}_\nu(p,q,-p-q)$ becomes bare for vanishing 
ghost momentum $p$~\cite{Taylor:ff},
\be
\lim _{p\to 0} \Gamma^{c\bar cA;abc}_\nu(p,q,-p-q) =  i g_3 f^{abc} q_\nu .
\label{taylor}
\ee
This identity especially ensures that we can choose a bare ghost-gluon vertex 
without altering qualitative results for the infrared behavior of the
propagators. Indeed, at least in four dimensions, the qualitative nature of the
infrared solution is independent of the detailed structure of the ghost-gluon
vertex to a large extent~\cite{Lerche:2002ep}. Numerical consistency checks also find 
that the ghost-gluon-vertex does not depart significantly from its tree-level 
form~\cite{Schleifenbaum:diploma}. Thus we will use a bare 
ghost-gluon vertex in the following. To complete the description of
our truncation we still have to choose the 3-gluon vertex and the Higgs-gluon
vertex. As these choices are motivated by more technical considerations we
will defer all details concerning these vertex functions to the appropriate 
sections describing the numerical solutions.

To obtain equations for the dressing functions, the color indices are 
contracted assuming the tree-level color structure for the vertex functions. This
assumption is supported by lattice calculations~\cite{Boucaud:1998xi}. The
gluon DSE is transformed to a scalar equation by contracting it with the tensor
(\ref{bpproj}). Dividing the DSEs by all trivial factors appearing on
the l.h.s.\ these equations read
\bea
{G^{-1}(p^2)}&=&1\nonumber\\
&+&\frac{g_3^2C_A}{4\pi^2}\int dq d\theta 
A_T(p,q)G(q^2)Z((p-q)^2)\; ,\nonumber\\ \label{fulleqG}\\
{H^{-1}(p^2)}&=&1+\frac{m_h^2}{p^2}+T^{HG}+T^{HH}\nonumber\\
&+&\frac{g_3^2C_A}{4\pi^2}\int dq d\theta \Big(N_1(p,q)H(q^2)Z((p+q)^2)
\nonumber\\
&& \qquad +N_2(p,q)H((p+q)^2)Z(q^2)\Big) \; ,\label{fulleqH}\\
{Z^{-1}(p^2)}&=&1+T^{GH}+T^{GG}\nonumber\\
&+&\frac{g_3^2C_A}{4\pi^2}\int  dq d\theta\Big(R(p,q)G(q^2)G((p+q)^2)\nonumber\\
&& \qquad +M_L(p,q)H(q^2)H((p+q)^2)\nonumber\\
&& \qquad +M_T(p,q)Z(q^2)Z((p+q)^2)\Big) \; , \label{fulleqZ}
\eea
where $\theta$ is the angle between $p$ and $q$. For the truncation used
in the present investigation, the integral kernels $A_T$,
$N_1$, $N_2$, $R$, $M_L$ and $M_T$ are given in 
appendix~\ref{appkernels}.\footnote{For convenience, there have been slight
redefinitions and changes of notation as compared to~\cite{Maas:2002if}.} 

\subsection{Constraints on the solutions}\label{Truncation}

Due to the truncation of the DSEs, the Slavnov-Taylor identities are not
fulfilled. This violation manifests itself in the appearance of spurious
divergences in the gluon self-energy. These can be removed by adjusting either
the tadpole terms or by projecting the gluon equation with the tensor
(\ref{bpproj}) and choosing $\zeta=d=3$ in the equation for the 3d transverse 
gluon propagator~\cite{Brown:1988bm}. The violation of gauge invariance  
does not  manifest itself in the Higgs equation
(\ref{higgs}) of the 3d reduced theory. Thus this equation is independent of
$\xi$. Furthermore, the Higgs self-energy in 3d is finite since the gluon 
propagator has only a logarithmic divergence in four dimensions. Therefore the
divergencies of the two tadpole terms in (\ref{higgs}) have to cancel each
other exactly. Nonetheless, as the Higgs has a finite mass, the tadpoles
involving a Higgs propagator have a finite contribution which correspond to a
finite mass renormalization. This has to be taken into account and will be
discussed in detail in section \ref{Analytic}.

Varying $\zeta$ while selecting the tadpoles to cancel any spurious
divergencies in the Yang-Mills and the Higgs sector of the gluon equation
(\ref{gluon}) allows to explore different degrees of gauge violations.
Especially the physical case of a transverse projected equation is of interest.
If the results do not depend qualitatively and only weakly quantitatively on
the value of $\zeta$ it is justified to assume that the effect of the violation
of gauge symmetry is small and under control, {\it c.f.} refs.\
\cite{Fischer:2002hn}. 

There is another aspect of gauge invariance to be respected in non-perturbative
calculations. As the Lorentz condition does not completely fix the gauge
field, configurations are over-counted. This is known as the Gribov problem~\cite{Gribov:1977wm}. 
The region of gauge space which contains no copies and includes the origin is called 
the fundamental modular region, see {\it e.g.}~\cite{Zwanziger:1993dh}. There is, 
however, at present no local condition known which defines this region. It is 
contained in the first Gribov region including the origin and configurations for
which the Faddeev-Popov determinant does not change sign. The restriction to the 
first Gribov region can be formulated at the  level of dressing functions as boundary 
conditions to the solutions of the DSEs as
\be
G(q^2)\ge 0,\quad Z(q^2)\ge 0,\quad H(q^2)\ge 0.\label{gribov}
\ee
The condition on $H$ does not follow from the Gribov condition in three
dimensions but only due to the fact that it is a component of the gauge field
in four dimensions. The condition (\ref{gribov}) does not completely solve the 
problem since the first Gribov horizon encloses a larger part of gauge space
than the fundamental modular region, and therefore contains gauge copies~\cite{vanBaal:1997gu}. 
Nevertheless, for the purpose of calculating propagators  it is likely that this 
condition is sufficient for eliminating Gribov copies~\cite{Zwanziger:2003cf}. 
Note that, in lattice calculations, the Gribov ambiguity poses a more severe
problem and makes it especially hard to extract the ghost propagator~\cite{Cucchieri:1997ns}.

\section{Infrared and ultraviolet behavior of the propagators}\label{Analytic}

\subsection{Ultraviolet behavior}\label{Ultraviolet}

The 3d reduced theory is also asymptotically free~\cite{Feynman:1981ss}. Thus
the propagators and vertex functions reduce to their tree-level values for
sufficiently large momenta. As the 3d reduced theory is finite, corrections are
power-like. (In the case of the Higgs this is only true for $k\gg m_h$.) Since
$g_3^2$ has the dimension of a mass and does not enter in the loop integrals 
of eqs.\ (\ref{fulleqG}), (\ref{fulleqH}) and (\ref{fulleqZ}), 
already by dimensional analysis, one infers that the one-loop
contributions are proportional to $g_3^2/q$. Thus all loop integrals are
subleading in the ultraviolet compared to the tree-level contribution. An
explicit verification of this behavior is given in appendix \ref{appUV}. 
Leaving aside the possibility of a finite renormalization, one has 
\be
\lim_{q^2\to\infty} G(q^2) =\lim_{q^2\to\infty} Z(q^2) = 
\lim_{q^2\to\infty} H(q^2)=1 .\label{uvtree}
\ee
For large enough $q^2$ the dressing functions are thus approximately given by 
\be
1+\frac{c_ig_3^2}{\sqrt{q^2}}\label{uvnlo}
\ee
where the constants $c_i$ turn out to be positive for all dressing functions. 
Hence all dressing functions approach the tree-level behavior from above.

\subsection{Infrared behavior}\label{Infrared}

In this subsection we will study the infrared behavior of the dressing functions
employing the following general ans{\"a}tze:
\bea
G(q)&=& A_g q^{-2g},\\
Z(q)&=& A_z q^{-2t},\label{gluoniransatz}\\
H(q)&=& A_h q^{-2l}.\label{iransatz}
\eea
For a pure Yang-Mills theory (without a Higgs) and a transversal projected
gluon DSE this analysis has already been performed for three space-time dimensions 
in ref.\ \cite{Zwanziger:2001kw}. 

The assumption $g=t=0$ and $l=-1$, motivated by the naive idea of a free gas of
quarks and gluons in the high-temperature limit, immediately leads to a
contradiction due to divergencies in the ghost and the gluon DSEs. Thus no such
'Coulomb phase' can be realized when taking into account condition
(\ref{taylor}) on the ghost-gluon vertex. Also $g=0$ and $t=l=-1$,
corresponding to a purely perturbative quark-gluon plasma with screening masses
in the magnetic and the electric sector, immediately leads to a contradiction.
This is also understood from the fact that, in order to obtain such a behavior,
the tadpoles would have to dominate. However, for $\zeta=3$, the tadpole in the 
gluon equation does not contribute. Besides, it is hard to see how the vacuum
corresponding to such propagators avoids the violation of Elitzur's 
theorem~\cite{Elitzur:im} (for further details see ref.\ \cite{Maas:phd}).
Hence the tadpole cannot be the infrared leading term in the gluon equation 
and one must have $t\le -1$. Convergence of the integrals in the DSEs even 
requires  $t<-1$, see appendix \ref{appIR}. This is in agreement with the 
reasoning~\cite{Zwanziger:2003cf} for infrared dominance of the gauge fixing 
part of the Lagrangian (\ref{lagrange}). The inequality
\be
t<-1 \label{tassump}
\ee
implies condition (\ref{Oehme}) and thus gluon confinement. On the one hand,
this is no surprise as one expects confinement also in the three-dimensional
Yang-Mills theory. On the other hand, it means that real-world chromomagnetic 
gluons are confined even at infinitely high temperatures!
Note that the solution found in ref.\ \cite{Zwanziger:2001kw} also obeys 
the inequality (\ref{tassump}).

In appendix \ref{appIR} the infrared limit of the DSEs is derived: 
\bea
\frac{y^g}{A_g}&=&y^{-(d-4)/2-g-t}I_{GT}(g,t)A_gA_z\label{ghostir}\\
\frac{y^t}{A_z}&=&1+y^{-(d-4)/2-2g}I_{GG}(g,\zeta)A_g^2\label{gluonir}\\
\frac{y^l}{A_h}&=&1+\frac{m_h^2+\delta m^2}{y}\label{higgsir}. 
\eea 
where $y=q^2$ and $d$ denote the dimension. A subtraction in (\ref{ghostir}) has 
been performed and only the finite part is retained. The expressions for
$I_{GT}$ and $I_{GG}$, originating from the ghost-self energy and the ghost-loop,
can be found in appendix \ref{appIR}. In the Higgs equation, a finite
renormalization of the mass has been allowed for. The mass renormalization is given in equation (\ref{higgstadpole}) and is discussed in \cite{Maas:phd}.
Eq.\ (\ref{higgsir}) possesses only one solution,  namely $l=-1$ and 
\be
A_h=\frac{1}{m_h^2+\delta m^2}.\label{ah}
\ee 
This indicates firstly a qualitative change in the high-temperature limit (as in
four dimensions at $T=0$ one has $t=l$) and secondly the decoupling of the Higgs
in the infrared. These observations agree with corresponding findings on the
lattice~\cite{Cucchieri:2001tw}.

The infrared ghost equation (\ref{ghostir}) implies the relation 
(see also~\cite{Zwanziger:2001kw})
\be
g=-\frac{1}{2}(t+\frac{4-d}{2})=_{d\to 3} \; \; 
-\frac{1}{2}(t+\frac{1}{2})\label{gt}.
\ee
The difference to the relation in four dimensions 
\be
g=_{d\to 4}\; \; -t/2\label{gt4d}
\ee
implies an additional power in momentum which exactly compensates the 
dimension of the effective coupling constant in three dimensions. A
corresponding compensation is expected at high temperatures in four dimensions
to cancel the straightforward temperature factor in front of the Matsubara sum.

Due to the inequality (\ref{tassump}) $g$ is positive. Thus any solution 
automatically satisfies (\ref{Kugo}) and both, the Kugo-Ojima and the 
Zwanziger-Gribov conditions, are fulfilled.

\begin{figure}
\epsfig{file=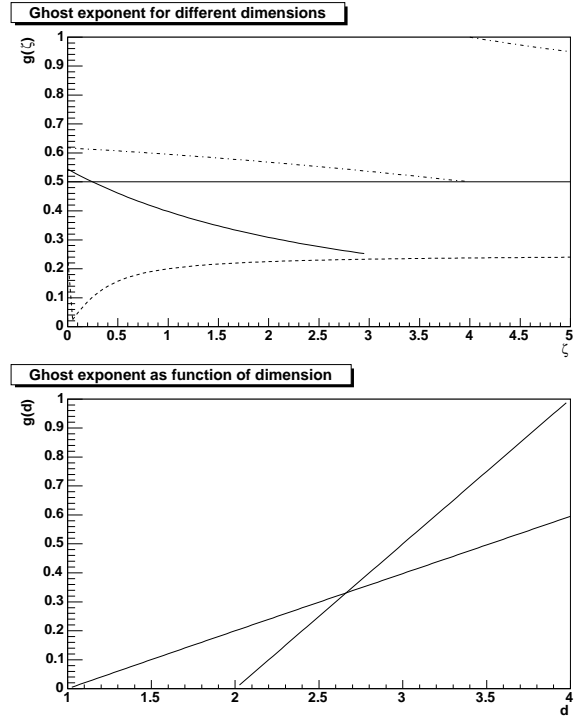,width=0.95\linewidth}
\caption{The upper panel shows the two solutions for the ghost exponent $g$ for
values of $\zeta$  allowed by condition (\ref{gribov}) and integral convergence, 
see Eq.\ (\ref{dglimits}) and ref.~\cite{Zwanziger:2001kw}. This excludes the 
second branch in $d=2$. 
The dashed line is for $d=2$, the solid line for $d=3$ and the dashed-dotted line
for $d=4$. 
The lower panel displays the two solutions for the ghost exponent $g$
as a function of $d$ for $\zeta=1$.}\label{figexp}
\end{figure}

Using (\ref{gt}) in Eq.\ (\ref{gluonir}) one obtains
\bea
-\Big(2^{2g-d}\sqrt{\pi}(\frac{d}{2}+g)(2+d(\zeta-2)-4g(\zeta-1)-\zeta) &
\nonumber\\
\times \csc(\frac{\pi(d-4g)}{2})\sin(\pi g)\Gamma(\frac{d}{2}+g)\Big)\Big/ &
\nonumber\\
\Big/\Big((d-1)^2g\Gamma(\frac{1+d-2g}{2})\Gamma(2g)\Big) &
\nonumber\\
=1.&
\label{irconsistanyd}
\eea
This equation has at least one solution for $d\ge1$ and two solutions for $d\ge
2$, see Fig.~\ref{figexp}. Eq.\ (\ref{irconsistanyd}) simplifies for $d=3$ to
\be
1=\frac{32g(1-g)(1-\cot^2(g\pi))}{(1+2g)(3+2g)(2+2g(\zeta-1)-\zeta)}.\label{irconsist}
\ee
One of the two solutions in $d=3$ is independent of the projection of the gluon
DSE:
\be
(g,t)=(\frac{1}{2},-\frac{3}{2})\label{exponents}
\ee
while the other one depends on the parameter $\zeta$. In the special case of a
transverse projection, $\zeta=1$, one has
\be
(g,t)=(0.3976,-1.2952)\label{sexponents}
\ee
thus reproducing the results of ref.\ \cite{Zwanziger:2001kw} where only the case 
$\zeta=1$ has been treated.

Condition (\ref{gribov}), the inequality (\ref{tassump}) as well as
convergence of the integrals in the infrared restrict $g$ in three dimensions
to
\be
\frac{1}{4}<g\le\frac{3}{4}.\label{gallowed}
\ee
Requiring furthermore a well-defined Fourier transform of the ghost
propagator (at least in the sense of a distribution) leads to the condition
$g\le1/2$. This restricts the range of allowed $\zeta$-values for the varying
branch to
\be
1/4\le \zeta<3.\label{zetaallowed}
\ee
At the lower boundary both solutions merge into one. 

Since in Eqs.\ (\ref{ghostir}) and (\ref{gluonir}) the  prefactors $A_g$ and
$A_z$ only appear in the product $A_g^2A_z$ only this combination is determined 
by the infrared analysis: 
\bea
\frac{1}{A_zA_g^2}&=&\frac{C_Ag_3^2}{(4\pi)^\frac{3}{2}}\nonumber\\
&&\times\frac{2^{4(g-1)}(2+2g(\zeta-1)-\zeta)\Gamma(2-2g)\sin^2(\pi g)}
{\cos(2\pi g)(g-1)g^2\Gamma(\frac{3}{2}-2g)}.
\nonumber\\ \label{ag}
\eea
Determining these pre-factors turns out to be an essential and unexpectedly 
complicated part of the numerical method. Thus Eq.\ (\ref{ag}) is used to check
whether a correct numerical solution is found~\cite{Maas:ccp}.

\section{Numerical Results}\label{Numerical}

The numerical solution of the DSEs (\ref{fulleqG}), (\ref{fulleqH}) and
(\ref{fulleqZ}) will be achieved in three steps: First, we study the ghost-loop
only truncation where only diagrams with at least one ghost propagator are
kept. Second, we include the gluon loop in the gluon equation. This truncation
scheme neglects all Higgs contributions and thus corresponds to a purely
three-dimensional Yang-Mills theory.  The last step is then to fully implement
the complete system. A description of the numerical method is given in refs.\
\cite{Maas:phd,Maas:ccp}.

\subsection{Ghost-loop-only truncation}\label{Ghost-only}

\begin{figure}
\epsfig{file=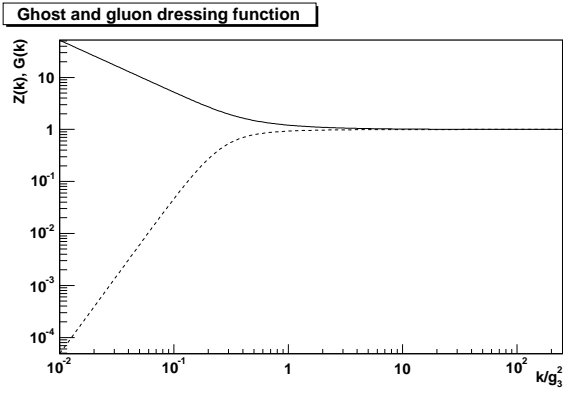,width=0.95\linewidth}
\caption{The solutions of the DSEs in the ghost-loop-only truncation scheme 
at $\zeta=3$. The dashed line refers to the dressing functions of the gluon $Z$ 
and the solid line to that of the ghost $G$. All quantities have been made 
dimensionless by dividing out appropriate powers of $g_3^2$.}\label{figgonly3}
\end{figure}

Keeping only ghost loops and using a bare ghost-gluon vertex, the system of DSEs
is completely specified.  The results of the corresponding calculation with
$\zeta=3$, for both the ghost  and the gluon  dressing functions, are shown in
Fig.~\ref{figgonly3}.  In this case, only the $g=1/2$ infrared solution
exists. The gluon propagator exhibits a  maximum located at $k/g_3^2\approx 0.25$.  

The coupling constant is the only dimensionful quantity entering the DSEs. 
Thus all results can be represented through dimensionless variables when
expressed in appropriate powers of $g_3^2$. Due to this fact, one can infer from 
Fig.~\ref{figgonly3} the results of the ghost-loop-only truncation for any
positive value of the coupling constant. Even for a very small coupling
constants strong non-perturbative effects for momenta smaller than
$g_3^2$ are clearly visible. Therefore, the infrared behavior of the pertinent 
2-point functions is never perturbative. 

\begin{figure}
\epsfig{file=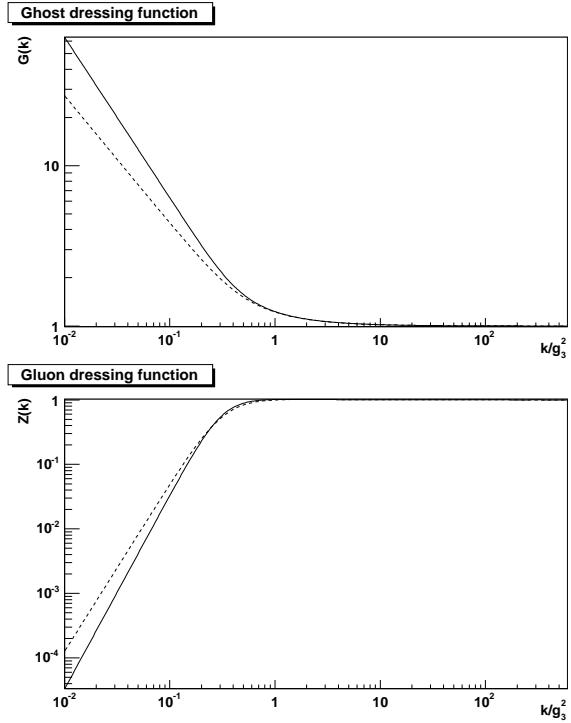,width=0.95\linewidth}
\caption{The ghost and gluon dressing function at $\zeta=1$. 
The upper panel shows the ghost dressing function, the lower panel shows the 
gluon dressing function. The solid line denotes the solution for 
$g=1/2$ and the dashed line gives the other solution branch with $g \approx 0.4$.}
\label{gonlyzeta1}
\end{figure}

For $\zeta\neq 3$, a spurious divergence appears in the ghost loop of the gluon
DSE. To cancel it, an appropriately adjusted tadpole term $T^{GG}$ is included.
A subtlety required here comes from the fact that the subtraction due to the tadpole 
is only effective in the ultraviolet and not in the infrared. A detailed account of
this subtraction is given in ref.\ \cite{Maas:phd} and the tadpole term $T^{GG}$
is given in appendix \ref{appTadpoles}.
The solutions for the transverse projection ($\zeta=1$) are displayed in
Fig.~\ref{gonlyzeta1}. In this case, solutions for both sets of infrared
exponents exist. As in the case $\zeta = 3$, the solutions show no special
features.

\begin{figure}
\epsfig{file=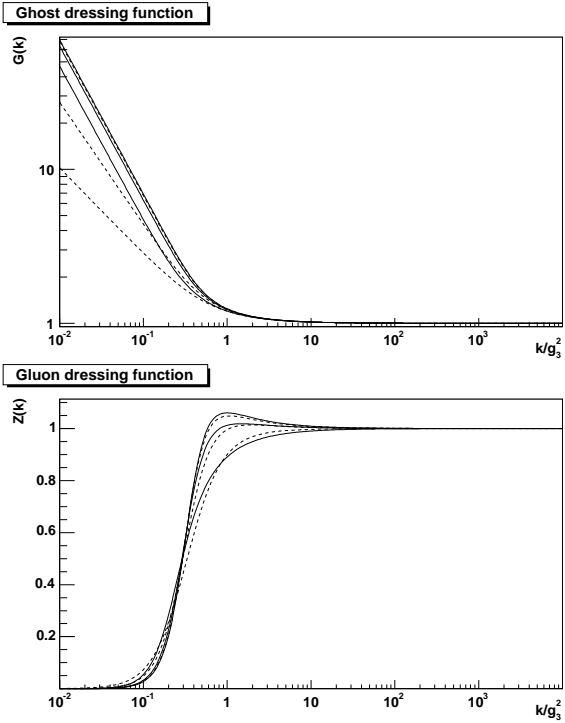,width=0.95\linewidth}
\caption{The ghost and gluon dressing function in ghost-loop-only approximation 
for different values of $\zeta$.  The upper panel shows the ghost dressing 
function, the lower panel the gluon dressing function. The solid lines represent 
the solutions with a ghost infrared exponent, $g=1/2$, and the dashed lines the 
ones for the other set of infrared exponents. At the peak in the gluon, the middle 
lines correspond to solutions at $\zeta=1$. The upper and lower line at mid-momenta 
give the solutions at $\zeta=0$ and $4$ for the $g=1/2$-branch and at $\zeta=0.25$ 
and $2.55$ for the other branch.}
\label{figgonlyzeta}
\end{figure}

To estimate the amount of gauge symmetry violation, the DSEs have been solved 
for different values of $\zeta$, see Fig.~\ref{figgonlyzeta}. To this end we
have varied $\zeta$ between 0 and 4 for the set of half-integer infrared exponents.
For the other branch, due to numerical uncertainties, $\zeta$ has only been
varied from 0.25 to 2.55, the later corresponding to $g=0.2708$ and
$t=-1.042$. Within the explored ranges the dependence of the solutions on 
$\zeta$, {\it i.e.\/} on the projection of the gluon DSE, is reasonably weak.

\subsection{Yang-Mills theory}\label{Yang-Mills}

\begin{figure}
\epsfig{file=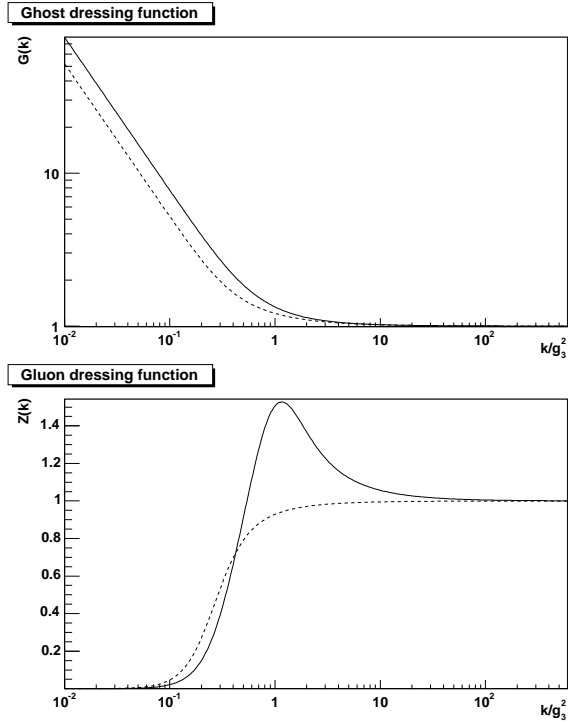,width=0.95\linewidth}
\caption{Solution of the Yang-Mills sector at $\zeta=3$. The dashed curve gives
the ghost-loop-only solution for comparison. The solid curve gives the full
Yang-Mills solution at $\delta=1/4$. Note the linear scale for $Z$ in the lower panel.}
\label{figym}
\end{figure}

\begin{figure}
\epsfig{file=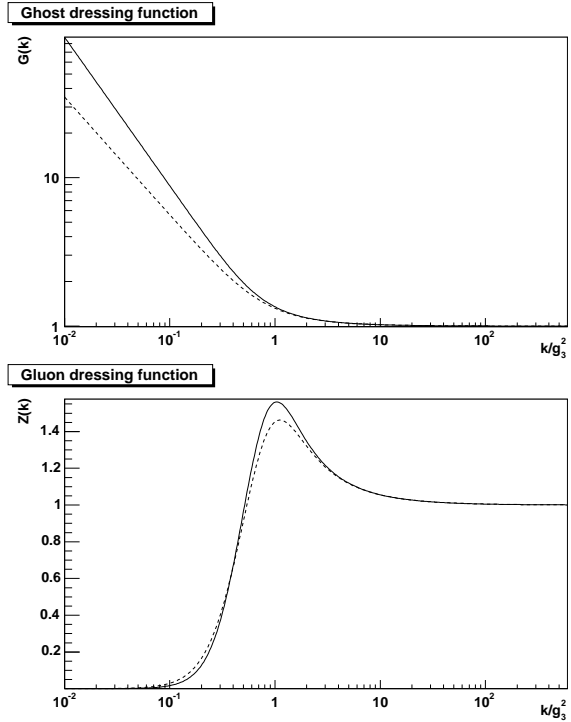,width=0.95\linewidth}
\caption{Solution of the Yang-Mills sector at $\zeta=1$. The upper panel shows
the ghost dressing function and the lower panel the gluon dressing function. The 
solid curve denotes the $g=1/2$ solution while the dashed curve displays the
$g\approx 0.4$ solution. Both are at $\delta=1/4$.}\label{figymzeta1}
\end{figure}

\begin{figure}
\epsfig{file=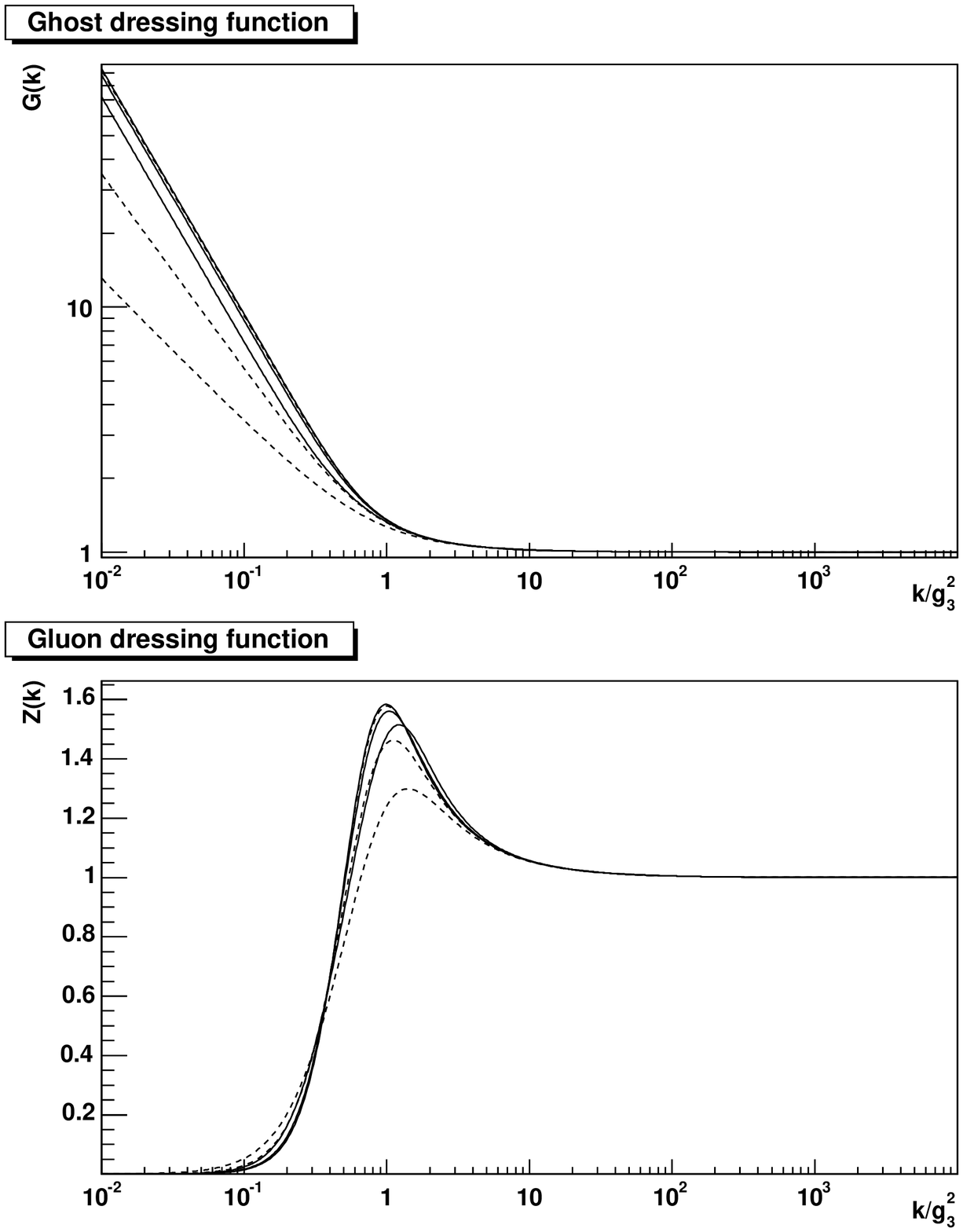,width=0.95\linewidth}
\caption{The ghost and gluon dressing of the pure Yang-Mills theory for
different values of $\zeta$. The upper panel shows the ghost dressing function, 
and the lower panel the gluon dressing function. The solid line gives the solution 
for $g=1/2$ and the dashed line is for the other solution branch. At the peak in 
the gluon dressing function, the middle lines represent the solution at $\zeta=1$. 
The upper and lower lines at mid-momenta give the solutions at $\zeta=0$ and
$\zeta=4$ for the $g=1/2$-branch and at $\zeta=1/4$ and $\zeta=2.45$ 
for the other solution branch.}
\label{figymzeta}
\end{figure}

\begin{figure}
\epsfig{file=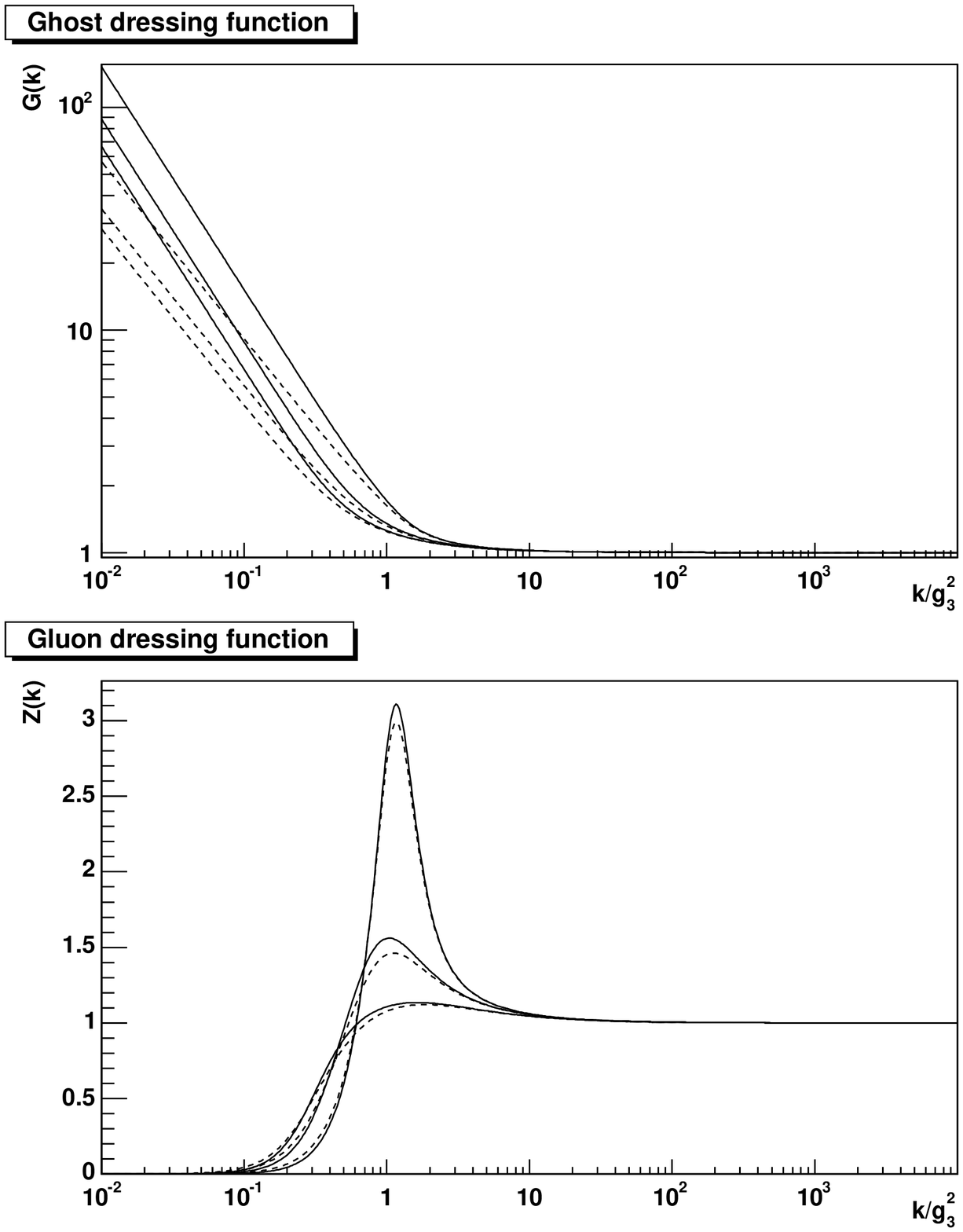,width=0.95\linewidth}
\caption{The ghost and gluon dressing functions of the pure Yang-Mills theory
for different values  of $\delta$. 
The upper panel shows the ghost dressing function, and the lower
panel the gluon dressing function. The solid line gives the
solution for $g=1/2$ and the dashed line is the other solution branch. At the peak
in the gluon dressing function, the middle lines represent the solution at
$\delta=1/4$. The upper and lower lines at mid-momenta give the solutions at
$\delta=0.131$ and $\delta=1$ for the $g=1/2$-branch and at $\delta=0.114$ and
$\delta=1$ for the other solution branch.} \label{figymdelta}
\end{figure}

The gluon loop with a bare 3-gluon vertex in the gluon DSE contributes with
opposite sign as the ghost-loop. This is correct in the ultraviolet and
irrelevant in the infrared where it is subleading. At momenta of the order of
$g_3^2$ the gluon loop calculated with a bare 3-gluon vertex becomes dominant.
This leads to zeros on the r.h.s.\ of the gluon DSE and thus to a violation of
the Gribov condition (\ref{gribov}). Therefore, within this truncation scheme,
the tree-level 3-gluon-vertex is not acceptable. Reasons might
be that at momenta around $g_3^2$ (i)~the bare 3-gluon-vertex overestimates the
true vertex, (ii)~the bare ghost-gluon vertex underestimates the true one
and/or (iii)~the two-loop graphs are important. These possibilities will be
explored in more detail in future work~\cite{Schleifenbaum:diploma}. However, 
as for the present study the
specific reason for this shortcoming is of minor importance, we proceed by
constructing an improved 3-gluon vertex. In this we are guided by earlier
studies in four dimensions~\cite{vonSmekal:1997is,Fischer:2003rp} where 
minimal modifications have been introduced to obtain the correct ultraviolet
behavior. To respect the Bose symmetry of the 3-gluon vertex we model
it by multiplying the bare 3-gluon vertex with an appropriate product 
of functions. This leads to the ansatz\footnote{Different ans\"atze, also ones
violating Bose symmetry, have been employed, and it has been found that they
do not lead to qualitative differences in the dressing function \cite{Maas:phd}.}
\bea
&&\Gamma^{A^3}_{\beta\sigma\mu}(-q,q+k,-k) \to \nonumber\\
&&\Gamma_{\beta\sigma\mu}^{tl;A^3}(-q,q+k,-k) \;
\left( A(q^2)A((q+k)^2)A(k^2)\right)^{-\delta} \; ,\nonumber\\
&&A(q^2)=Z(q^2)(G(q^2))^{2+1/2g} . 
\label{g3vertex}
\eea
The function $A$ is chosen to be a constant in the infrared.  The additional
parameter $\delta$ does not change this behavior. On the other hand, it allows
to smoothly interpolate between a large suppression and the tree-level value by
tuning it from large positive values to zero. Stable solutions have been found
for $\delta \ge 0.114$. In the following, we employ mostly $\delta=1/4$ which
yields a mild suppression of the 3-gluon vertex without effecting the stability
of the numerical calculation. 

The resulting dressing functions, using $\zeta=3$, are sho\-wn in comparison to
the ghost-loop-only result in Fig.~\ref{figym}.
For $\zeta\not=3$ tadpole terms have again to be subtracted. Their construction 
is detailed in ref.\ \cite{Maas:phd} and the corresponding expressions are 
given in appendix \ref{appTadpoles}. The results for $\zeta=1$ and $\delta=1/4$ are
shown in Fig.~\ref{figymzeta1}. The dependence on $\zeta$ which provides
a measure of gauge invariance violation, can be inferred from
Fig.~\ref{figymzeta} and the one on $\delta$ from Fig.~\ref{figymdelta}.
The only qualitative difference in these functions, when compared to the ones of
the ghost-loop-only truncation, is a maximum in the gluon dressing function 
which is now present for all employed values of $\zeta$. The location and height
of this maximum depends on the vertex construction. For most values of $\delta$ 
this dependence is weak, however. Nevertheless, we have to
conclude that the behavior of the dressing functions is sensitive to the
structure of the 3-gluon vertex and additional information about this
vertex function is highly desirable and will be studied in the future. 
In the context of this paper we will rely on the comparison to lattice data 
(see subsection \ref{Lattice}) to justify our ansatz.

Being confident that these technical issues do not invalidate our results 
we note that the infrared behavior of the gluon and ghost propagators in 
three-dimensional Yang-Mills theory is very similar to the ones within
four-dimensional Yang-Mills theory. Differences are of minor, quantitative
nature. This is not surprising since the three-dimensional Yang-Mills theory
is expected to be also a strongly interacting, confining theory~\cite{Feynman:1981ss}.

\subsection{Including the Higgs}\label{Higgs}

Finally, we add the Higgs, {\it i.e.\/} we implement the full system of
equations  (\ref{fulleqG}), (\ref{fulleqH}) and (\ref{fulleqZ}).  Although the
Higgs loop in the gluon DSE provides a positive contribution it is not
sufficient to allow for solutions with a bare 3-gluon vertex.  Thus, the ansatz
(\ref{g3vertex}) will also be used for the full system. On the other hand, the
tree-level gluon-Higgs vertex will be employed in this section. (In subsection
\ref{Masses} also a modified gluon-Higgs  vertex will be studied.)

As stated already, a tree level Higgs mass is induced when integrating out the
higher Matsubara modes in the process of dimensional reduction~\cite{Kajantie:1995dw}.  
For the current analysis, the origin and the exact
value of the Higgs mass are not of direct importance. Hence this mass will be
fixed to a value extracted from lattice calculations and we use 
$m_h/g_3^2=0.8808$~\cite{Cucchieri:2001tw} in the following.

As can be inferred from Fig.\ \ref{ftsys} three additional tadpole contributions
arise when the Higgs is included, see also appendix \ref{appdse}. Since the
Higgs has a tree-level mass, two of the tadpoles can already have a
non-vanishing finite part in leading-order perturbation theory, see appendix
\ref{appUV}. 
The self-energy of a Higgs field in a three-dimensional theory would be in 
general linearly divergent. This is not the case here. The Higgs, being a 
component of the gluon field in the four-dimensional theory, protects
its self-energy from being divergent at the expense of fixing at leading-order 
perturbation theory the tree-level coupling $h$ to be
\be
h=-2g_3^2\frac{C_A}{C_A+2},\label{hvalue}
\ee
see appendix \ref{appUV}. 

\begin{figure}
\epsfig{file=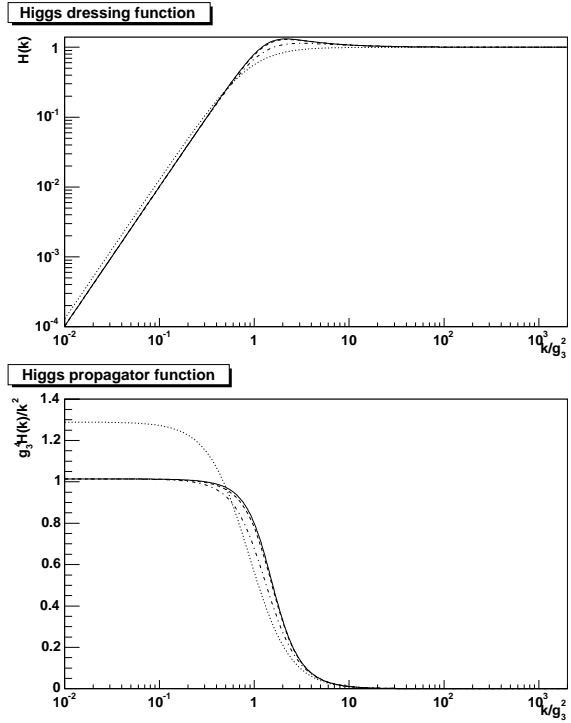,width=0.95\linewidth}
\caption{The Higgs dressing function and propagator at $\zeta=1$. The solid
line gives the solution for $g=1/2$, the dashed line for $g\approx 0.4$, the
dashed-dotted line denotes the leading-order perturbative result and the dotted line
the tree-level behavior.} \label{fighiggsf}
\end{figure}

As can be seen from Fig.\ \ref{figymf}, the influence of the Higgs on the
Yang-Mills sector is small. This is in agreement with results from
lattice calculations~\cite{Cucchieri:2001tw}. The resulting Higgs propagator,
see Fig.\ \ref{fighiggsf}, behaves similar to a massive tree-level 
propagator.

\begin{figure}
\epsfig{file=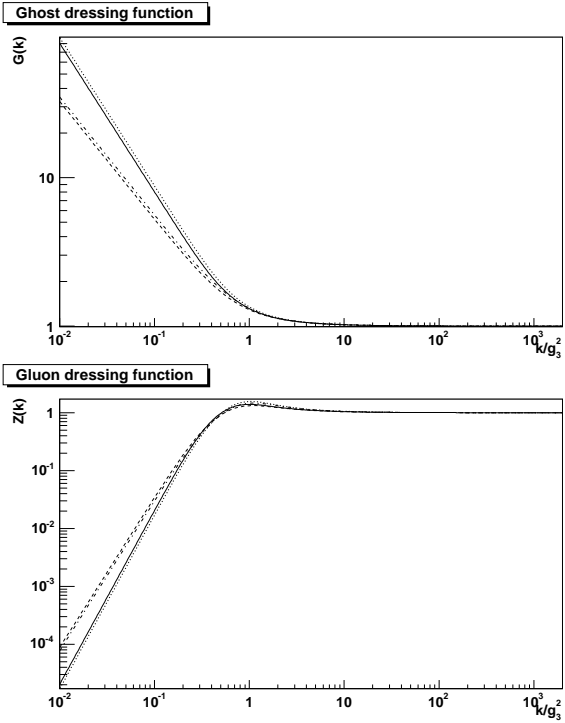,width=0.95\linewidth}
\caption{The solution of the Yang-Mills system compared to the full system at
$\zeta=1$. The solid line denotes the full solution at $g=1/2$ while the dotted
line is the corresponding Yang-Mills solution. The dashed line gives the full
solution for the $g\approx 0.4$ solution while the dashed-dotted line is the
corresponding Yang-Mills solution.} \label{figymf}
\end{figure}

Although the Higgs mass is fixed in the current setting to be 
$m_h=0.8808  g_3^2$, the Higgs mass dependence of
the gluon and ghost propagators, especially for decreasing values of the Higgs
mass,\footnote{As expected, for large values of the tree-level Higgs mass,
the solutions for the gluon and ghost propagators are indistinguishable from
the corresponding solutions of the pure Yang-Mills theory.} would be of interest.
However, already after a slight decrease in the Higgs mass, the solution
ceases to exist. For masses below $m_h\approx 0.6 g_3^2$ no solution is found 
(at least, for $\delta\le 1$).

\begin{figure}
\epsfig{file=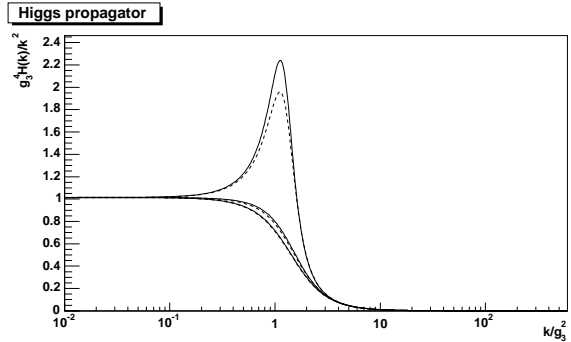,width=0.95\linewidth}
\caption{The Higgs propagator as a function of $\delta$
at $\zeta=1$. The solid line denotes the solution for $g=1/2$ and the dashed line for
$g\approx 0.4$. The higher the peak the smaller $\delta$. The lowest peak
corresponds to $\delta=1$, the middle one to $\delta=1/4$ and the largest one
to $\delta=0.0862$ and $\delta=0.061$, respectively.} \label{fighiggsdelta}
\end{figure}

The dependence of the Higgs on $\zeta$ is very small, deviations
between solutions for different values of $\zeta$ not exceeding more than a few 
percent. However, the Higgs dressing function, at momenta $k\approx
g_3^2$, is sensitive to the parameter $\delta$, see Fig.\ \ref{fighiggsdelta}. 
Similar to what happens in the gluon dressing function, also the maximum of the
Higgs dressing function increases with decreasing $\delta$. 
This is understandable from the fact that the Higgs self-energy depends
on the strength of the gluon propagator.

\begin{figure}
\epsfig{file=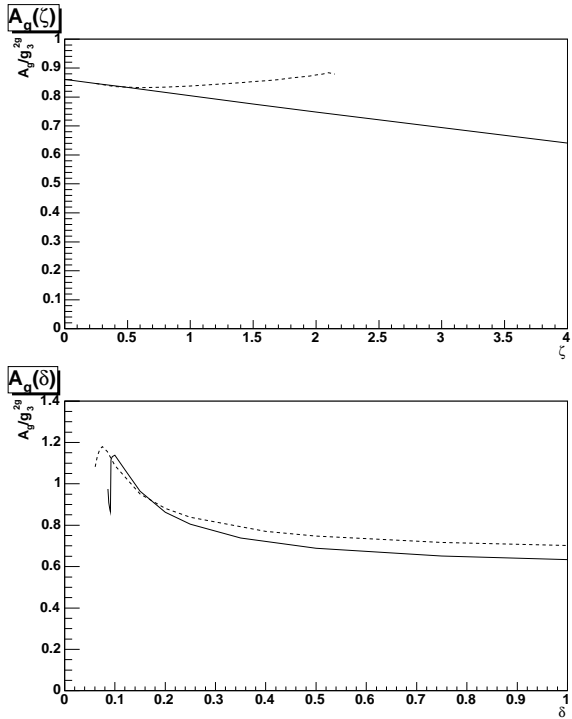,width=0.95\linewidth}
\caption{The dependence of the infrared coefficient $A_g$ on $\zeta$ in the
upper panel and on $\delta$ in the lower panel. Solid lines represent the $g=1/2$
solution and dashed lines the other solution branch.} \label{figagzd}
\end{figure}

Although the analysis reveals some dependence on the parameters
$\zeta$ and $\delta$ at intermediate momenta, it is important to note that
infrared properties are only weakly dependent on these superficial quantities.
Only one of the two exponents depends mildly on $\zeta$ as already discussed. 
In Fig.\ \ref{figagzd} we display the infrared
coefficient $A_g$ as function of $\zeta$ or $\delta$. (Note that $A_h$ is fixed 
by the renormalized mass and $A_z$ depends uniquely on $A_g$.) 
Besides some dependence on $\delta$ for small values of this parameter (where
the solutions cease to exist, as well) the infrared coefficients
are rather robust up to the edges of numerical stability.

\subsection{Comparison to lattice results}\label{Lattice}

Recently lattice results for the gluon and the Higgs propagator~\cite{Cucchieri:2001tw} 
have become available. As the gluon propagator shows almost
no sensitivity to the Higgs, not only in our but also in these lattice
calculations, we additionally  compare to the gluon propagator computed within
three-di\-mensional lattice Yang-Mills theory \cite{Cucchieri:2003di}. 

\begin{figure}
\epsfig{file=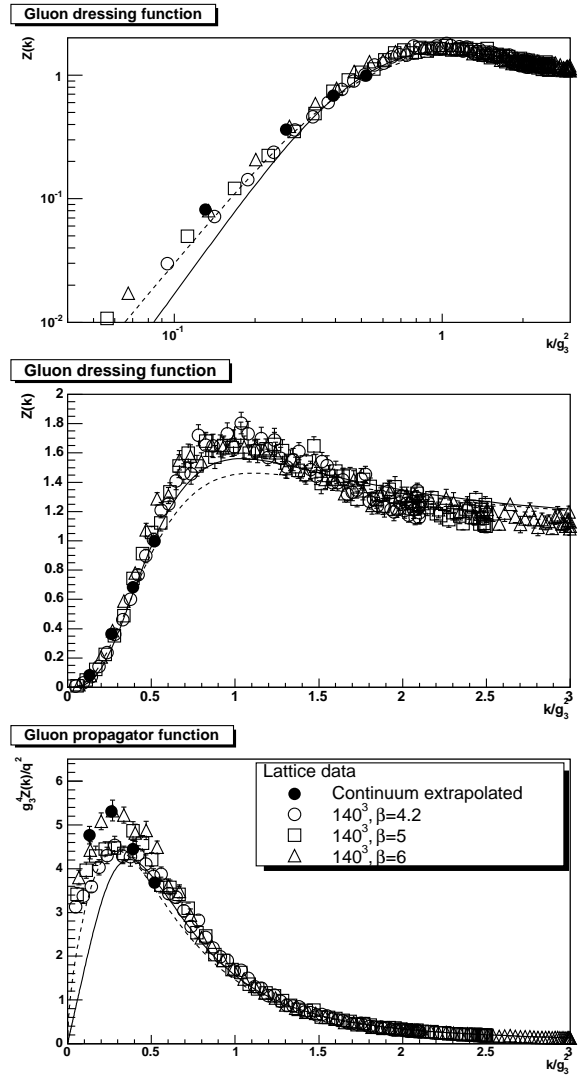,width=0.95\linewidth}
\caption{The gluon propagator and dressing function in pure Yang-Mills theory
from the lattice and from subsection \ref{Yang-Mills}. The
continuum-extrapolated values are from~\cite{Cucchieri:2001tw}, the others from
\cite{Cucchieri:2003di}. The errors indicated are statistical. The solid line
denotes the $g=1/2$ solution and the dashed line the $g\approx 0.4$ solution. Both are
at $\zeta=1$ and $\delta=1/4$.} \label{figymlat}
\end{figure}

\begin{figure}
\epsfig{file=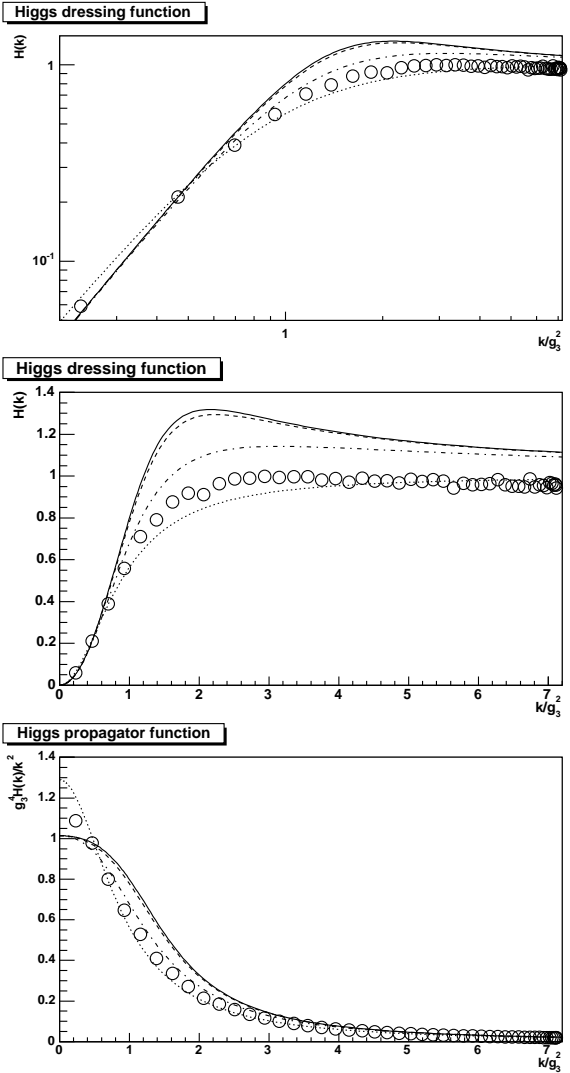,width=0.95\linewidth}
\caption{The Higgs propagator and dressing function in the full theory from the
lattice and from subsection \ref{Higgs}. The lattice data are from
\cite{Cucchieri:2001tw}. The errors indicated are statistical. The solid lines
give the $g=1/2$ solution and the dashed lines the $g\approx 0.4$ solution, both at
$\zeta=1$ and $\delta=1/4$. The dotted lines represent the tree-level result while 
the dashed-dotted lines give the leading-order perturbative results.}
\label{fighiggslat}
\end{figure}

As can be inferred from Fig.\ \ref{figymlat}, our results for the gluon propagator
agree astonishingly well with the corresponding lattice results. As for the
infrared behavior, the lattice results are in favor of the $g\approx 0.4$
solution. Note that, at momenta $k\approx g_3^2$, our truncation scheme is not
trustworthy and the better agreement of the $g=1/2$ solution with the lattice
gluon dressing function is not conclusive.
As displayed in Fig.\ \ref{fighiggslat} the DSE results for the Higgs propagator
show clear deviations from the corresponding lattice results, the latter being
closer to the leading-order perturbative results than to the DSE results.
We will return to a discussion of this point in subsection \ref{analytic}.

\section{Towards observables}\label{Derived}

In a first step towards a calculation of observables from the propagators we
determine an approximative form of the thermodynamic potential.
Furthermore, we investigate screening masses via the calculation of the
Schwinger functions from the propagators. In the following,
the results for $\zeta=1$, $\delta=1/4$ and $C_A=N_c=3$ will always be used.

\subsection{Thermodynamic potential}\label{td}

Knowledge of all Green's functions as functions of the temperature would allow
to calculate the thermodynamic potential and therefore all thermodynamic
quantities. Since not all of these are known, it is not possible to calculate
an exact thermodynamic potential. Knowledge of the propagators is however 
sufficient to compute an approximation to the thermodynamic potential, the
Luttinger-Ward  or Corn\-wall-Jackiw-Tomboulis (LW/CJT) effective action
\cite{Luttinger:1960ua}. 

This effective action can be extended to abelian and non-abelian gauge 
theories, see {\it e.g.\/}~\cite{Haeri:hi,Carrington:2003ut}. Its
calculation is, however, prohibitively complicated when using constructed instead 
of bare or exact vertices. Therefore we will only extract here some  qualitative
features using the simplest approximation to the thermodynamic potential. It
only depends on the propagators, and one has
\bea
\Omega=&\frac 1 2 d(G) T\sum\int\frac{d^3p}{(2\pi)^3}\Big(-\ln(Z(q))+(Z(q)-1)
\nonumber\\
&+\ln(G(q))-(G(q)-1)\nonumber\\
&-\frac{1}{2}\ln(\frac{H(q)}{H_0(q)})+\frac{1}{2}(\frac{H(q)}{H_0(q)}-1)\Big)
\label{cjt}.
\eea
where $d(G)=\delta^{aa}$ is the dimension of the gauge group. 
$H_0$ is the tree-level Higgs dressing function
\be
H_0(q)=\frac{q^2}{q^2+m_h^2} \; .
\ee
Working in the infinite-temperature limit we are eventually interested 
in the energy density divided by $T^4$. This motivates a rescaling of the 
integration momenta and one obtains 
\be
\frac{\Omega}{T^4}=\frac{g_3^6}{T^3}a\label{eps}
\ee
where the dimensionless constant $a$ depends only on the dimensionless ratio
$m_h/g_3^2$ and thus becomes independent of temperature in the limit 
$T\to\infty$. In the
simplest case, $g_3^2\sim g_4^2(\mu)T$ with the four dimensional coupling
constant $g_4$ depending on the renormalization scale $\mu$. Herein enters the
way in which the limit $T\to\infty$ is performed. When  
keeping $g_3(\mu)$ fixed by appropriately choosing $\mu(T)$, one has
$g_3^2\sim\Lambda_{QCD}$. Hence (\ref{eps}) scales as $1/T^3$ and
does not contribute to the thermodynamic pressure significantly compared
to the hard modes. The Stefan-Boltzman behavior must then be obtained from the
hard modes alone, requiring 
\be
\frac{\Omega}{T^4}=g^6_4(\mu)\bigl (a+\frac{b}{g_4^6(\mu)}\bigr ) = 
\frac{g_3^6}{T^3}a+ b
\ee
where $b$ stems from the hard modes~\cite{Maas:phd}. However, the soft modes
may still contribute significantly to thermodynamic properties, especially the
trace anomaly, near the phase transition~\cite{Zwanziger:2004np}.

The calculation of (\ref{cjt}) turns out to be plagued by spurious divergences,
since the DSEs with the modified 3-gluon-vertex are no longer exact stationary
solutions of (\ref{cjt}). After carefully subtracting these~\cite{Maas:phd},
$a$ turns out to be within errors of ${\cal O}(1)$, but significantly depending
on the truncation and large numerical uncertainties. The best value for the
full solution is 3.6 for the $g=1/2$ solution and 3.5 for the $g\approx 0.4$
solution, contributed mostly from the Higgs-sector. The difference of two
phases can be extracted with better accuracy, as the leading spurious
divergence cancels. The result is
\be
a_{g=1/2}-a_{g\approx 0.4}=0.053~.
\ee
Thus the $g\approx 0.4$ solution is thermodynamically preferred, in agreement with the
lattice results in subsection \ref{Lattice}.

\subsection{Schwinger functions}\label{Masses}

Screening masses are most directly extracted from the analytic structure of the
propagators. To obtain access to these analytic properties we calculate
the Schwinger function related to the propagator $D$, defined as \cite{Alkofer:2003jj}
\be
\Delta(t)=\frac{1}{\pi}\int_0^\infty dp_0\cos(tp_0)D(p_0),\label{schwinger}
\ee
{\it i.e.\/} the Fourier transform of the propagator with respect
to (Euclidean) time. Note that this definition is independent of the 
dimensionality of the underlying theory. Negative
values for the Schwinger function can be traced to violations of positivity 
and therefore to absence of the particle, represented by $D$, from the physical
spectrum~\cite{Alkofer:2003jj}. 

\begin{figure}
\epsfig{file=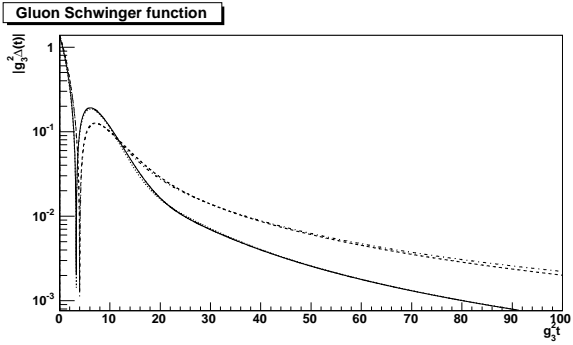,width=0.95\linewidth}
\caption{The Schwinger function of the gluon. The solid line gives the numerical
result for $g=1/2$ while the dashed line is for $g\approx0.4$. The dotted and dash-dotted
line denote their respective fits using the ansatz (\ref{gluonfit}).} 
\label{figzschwing}
\end{figure}

Using a sufficiently sophisticated FFT-algorithm together with at least 512 or
more frequencies\footnote{Better are several ten thousands to a million.  The
results presented here have been obtained using roughly $5\cdot10^5$ frequencies.}, it is possible to calculate the Schwinger functions.  The
result for the gluon, presented in Fig.\ \ref{figzschwing}, clearly exhibits
positivity violations. This is in accordance with the Oehme-Zimmer\-mann
super-convergence relation (\ref{Oehme}) and thus has been expected.
Furthermore, the position of the zeros can be interpreted as the confinement scale.
For the $g=1/2$ solution, the zero occurs at $g_3^2t\approx 3.29$  and at
$g_3^2t\approx 3.98$ for the other branch. This result is in agreement with recent lattice results~\cite{Cucchieri:2004mf}.

\begin{figure}
\epsfig{file=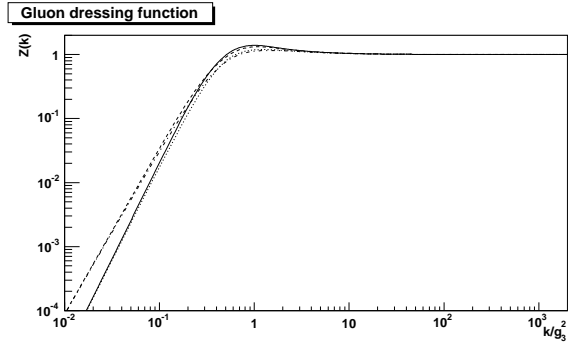,width=0.95\linewidth}
\caption{Comparison of the fit-function (\ref{gluonfit}) to the full solutions
of the dressing function. The solid line indicates the $g=1/2$ solution and dotted 
line its fit. The dashed line represents the $g\approx 0.4$ solution and the 
dashed-dotted line its fit.}\label{figgluonprop}
\end{figure}

To be able to perform an analytic continuation of the gluon propagator into
the complex $q^2$-plane we first fit the Schwinger function.  This is performed
by using the ansatz\footnote{In ref.\ \cite{Alkofer:2003jj} a parameterization
with only a branch cut and no isolated pole provided a successful fit. Due to
the different  asymptotic behavior in four and three dimensions we use a
different ansatz here.}
\bea
Z_f(q)&=\frac{A_z q^{4g+1}}{1+f+A_z q^{4g+1}}(1+\frac{f}{1+fa_uq})
\label{gluonfit}\\
&=\frac{A_z q^{4g+1}(1+f+a_ufq)}{(1+a_ufq)(1+f+A_zq^{4g+1})}\nonumber
\eea
\noindent for the dressing function. $A_z$ is the infrared coefficient determined previously, and $a_u$ is the
ultraviolet coefficient of leading-order resummed perturbation theory, as
calculated in appendix~\ref{appUV}. The fit parameters for both solutions are
given in table~\ref{tabglfit}. As demonstrated in Fig.\ \ref{figzschwing} the
Schwinger function is fitted very well. The gluon propagator and dressing
function are also reasonably well described by the fit, see Fig.\
\ref{figgluonprop}.

\begin{table}
\begin{center}
\begin{tabular}{|c|c|c|c|}
\hline
Solution & $A_zg_3^{-2t}$ & $a_ug_3^2$ & $f$ \cr
\hline
$g=1/2$ & 20.3 & $64/27$ & 1.32511 \cr
\hline
$g\approx 0.4$ & 13.4 & $64/27$ & 1.03148 \cr
\hline
\end{tabular}
\end{center}
\caption{The coefficients for the gluon fit (\ref{gluonfit}).}
\label{tabglfit}
\end{table}

\begin{figure}
\epsfig{file=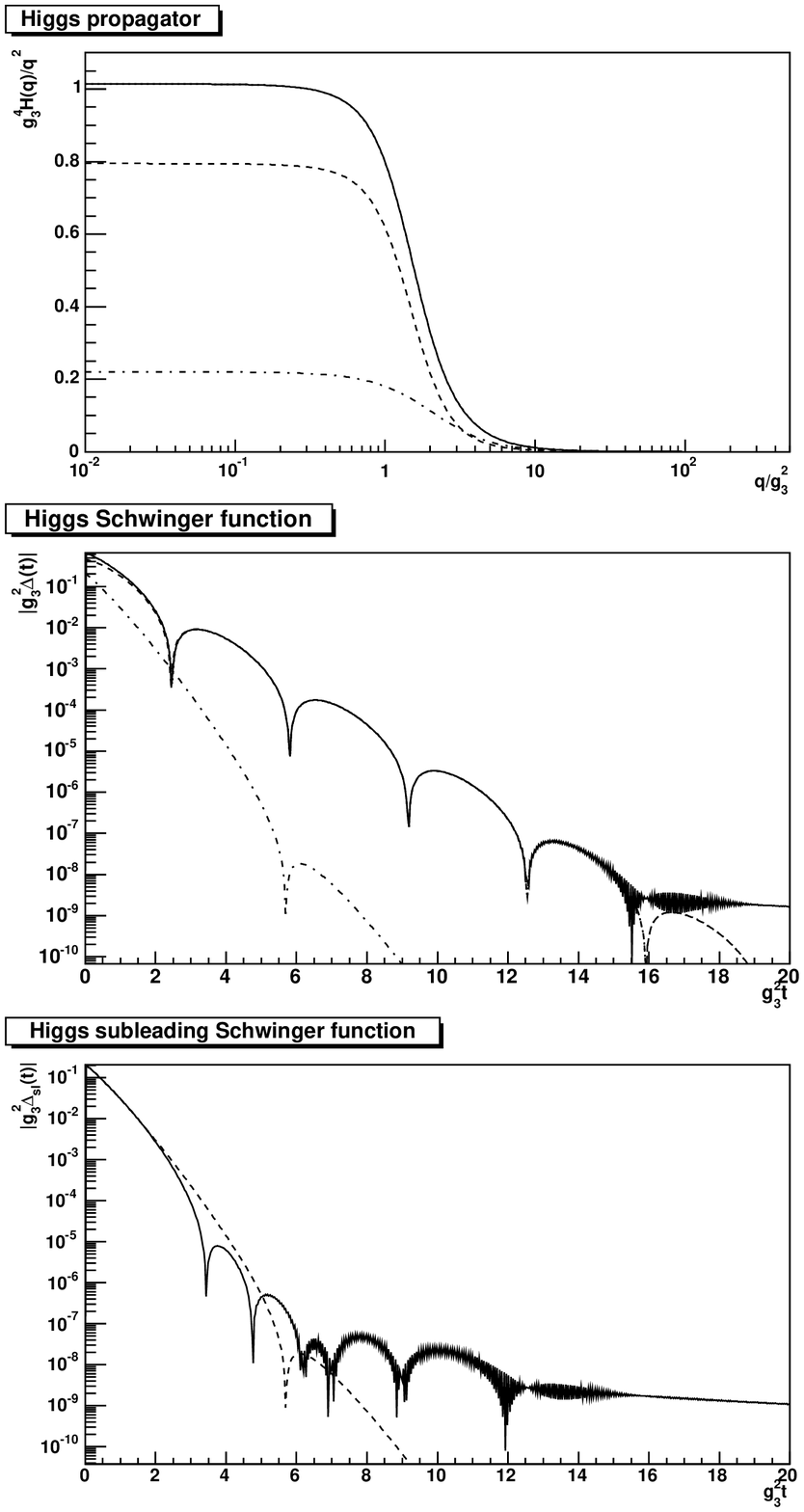,width=0.95\linewidth}
\caption{The Higgs propagator in the top panel and its Schwinger function in
the middle panel compared to their fits for $g=1/2$. The solid lines represent the
numerical solution, the dashed lines give the leading contribution and the
dashed-dotted lines the first subleading contribution. The dotted lines
underneath the solid lines give the sum of the leading and subleading contribution.
The bottom panel shows the comparison of the numerical (solid) and the fitted
(dashed) subleading Higgs contribution, see text.}
\label{fighstdfit}
\end{figure}

\begin{figure}
\epsfig{file=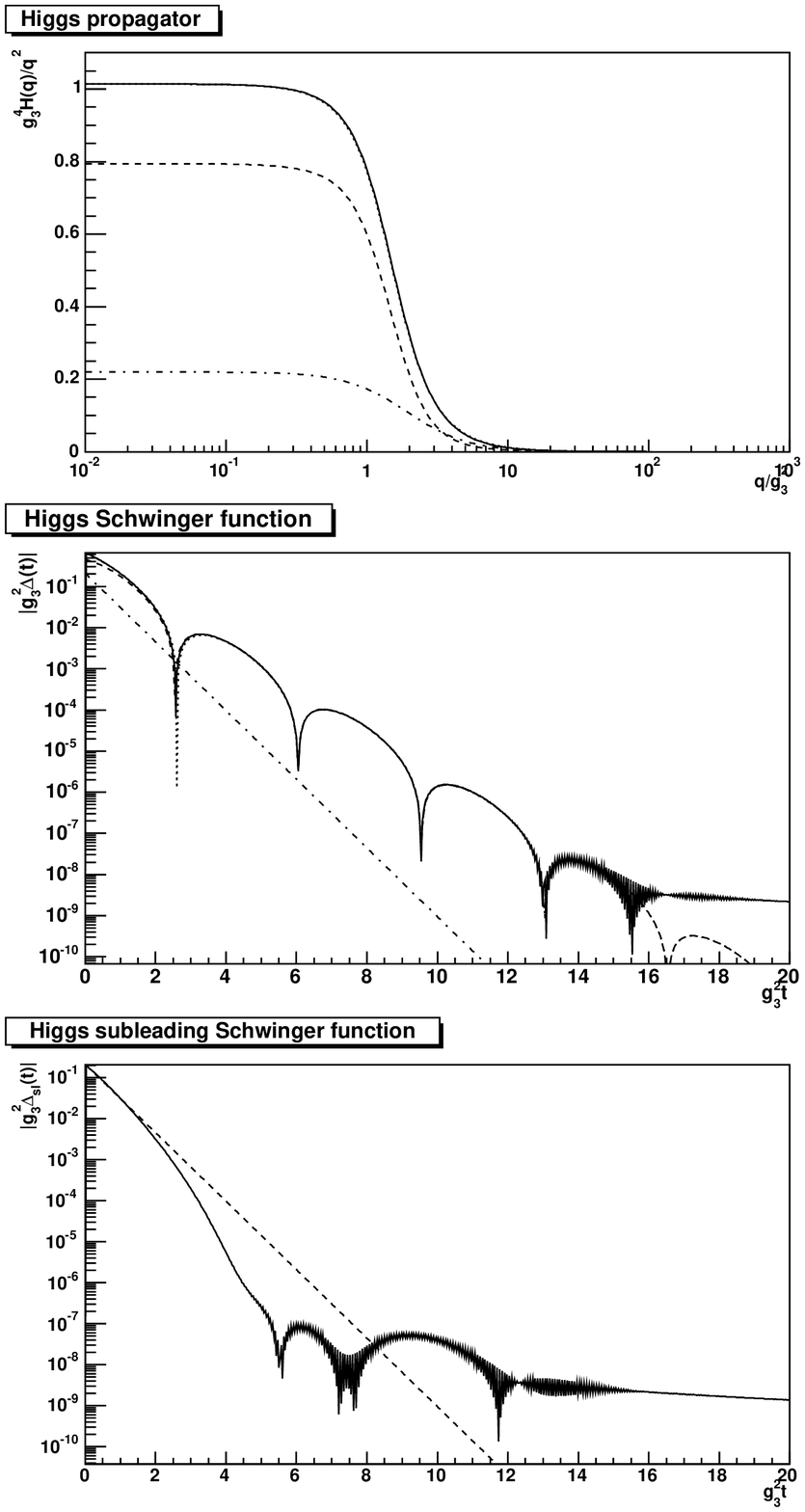,width=0.95\linewidth}
\caption{The Higgs propagator in the top panel and its Schwinger function in
the middle panel compared to their fits for $g\approx 0.4$. The solid lines represent the
numerical solution, the dashed lines give the leading contribution and the
dashed-dotted lines the first subleading contribution. The dotted lines
underneath the solid lines give the sum of the leading and subleading contribution.
The bottom panel shows the comparison of the numerical (solid) and the fitted
(dashed) subleading Higgs contribution, see text.}
\label{fighaltfit}
\end{figure}

Similar to a meromorphic fit ansatz used in ref.\ \cite{Alkofer:2003jj}
we will describe the Higgs propagator and its Schwinger function,
the latter being analytically calculated from the former, as follows
\bea
&H_f(q)=\frac{e+f q^2}{q^4+2m^2\cos(2\phi)q^2+m^4} \; ,\label{higgsfit}\\
&\Delta_f(t)=\frac{e}{2m^3\sin(2\phi)}e^{-tm\cos(\phi)}\nonumber\\
&\times(\sin(\phi+tm\sin(\phi))+\frac{fm^2}{e}\sin(\phi-tm\sin(\phi))) \; .
\nonumber
\eea
As demonstrated in Figs.\ \ref{fighstdfit} and \ref{fighaltfit} these fits 
already describe the
Schwinger function quite well, but miss around 20\% of the propagator at zero
momentum. This indicates that further massive modes are present. Indeed, for the
$g=1/2$ solution, a further term of the form (\ref{higgsfit}) has to be added. 
For the $g\approx 0.4$ solution, adding a term with one pole,
\bea
H_f(q)=\frac{e}{q^2+m^2} \; ,\label{higgsaltfit}\\
\Delta_f(t)=\frac{e}{2m}e^{-mt}\; ,\nonumber
\eea
allows to improve the fit in this case, as well. Both subleading fits are 
not very accurate for large $t$, and the results have to be taken with
care. They indicate, nevertheless, the existence of subleading contributions due
to the presence of further massive-particle-like contributions. The fit
parameters of both solutions can be found in table \ref{tabhfit}.

\begin{table}
\begin{center}
\begin{tabular}{|c|c|c|c|c|}
\hline
Solution & $e$ & $fg_3^4$ & $\phi$ & $m/g_3^2$ \cr
\hline
$g=1/2$ & 4.0199 & 0.3545 & -0.67078 & 1.4998 \cr
\hline
subleading & 9.4493 & 0.6736 & -0.18387 & 2.561 \cr
\hline
$g\approx 0.4$ & 4.0697 & 0.37793 & -0.63975 & 1.5045 \cr
\hline
subleading & 0.81188 & & & 1.9223 \cr
\hline
\end{tabular}
\end{center}
\caption{The coefficients for the Higgs fit (\ref{higgsfit}) and the subleading
one (\ref{higgsfit}) and (\ref{higgsaltfit}), respectively.} \label{tabhfit}
\end{table}

\begin{figure}
\epsfig{file=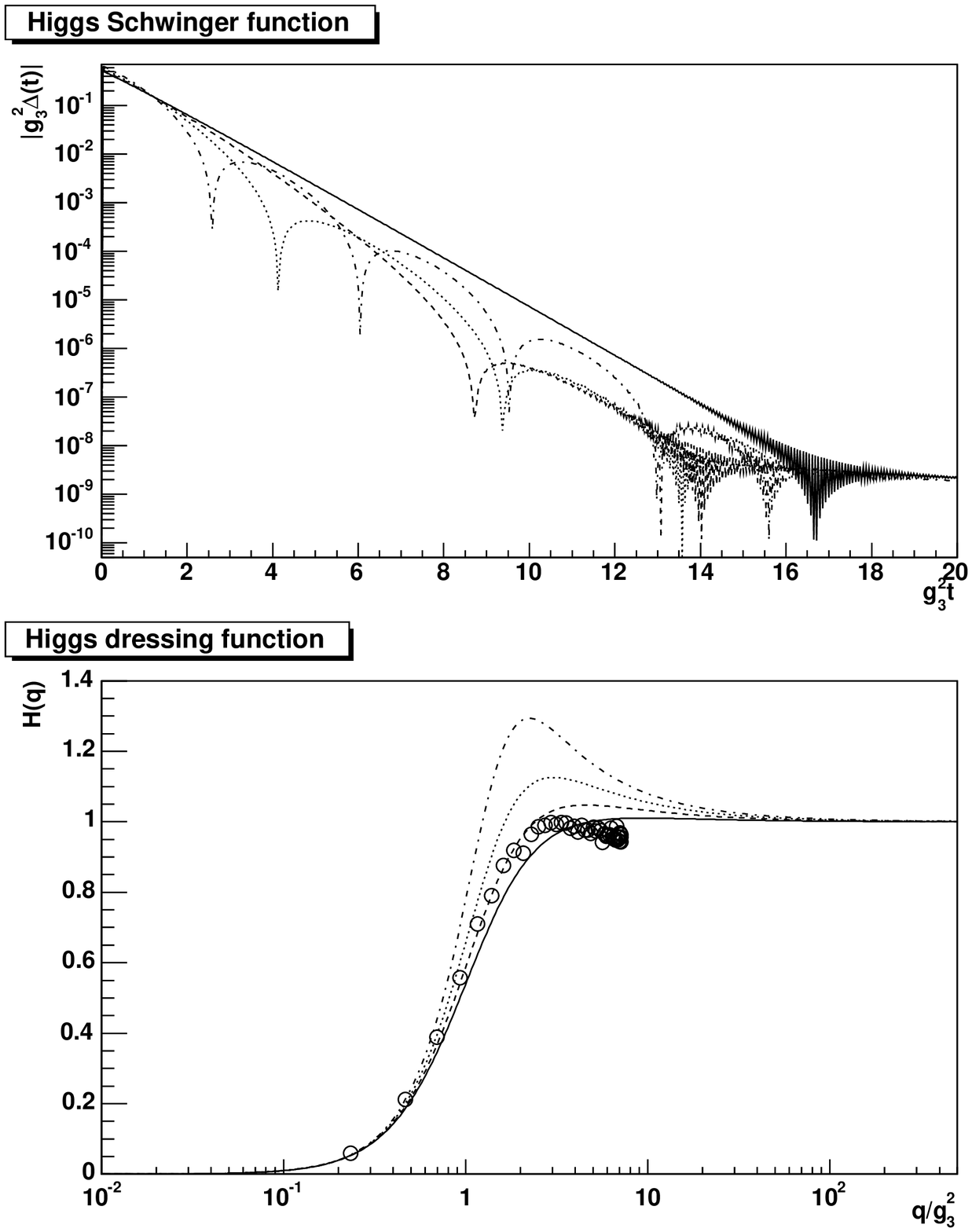,width=0.95\linewidth}
\caption{The top panel shows the Higgs propagator of the $g\approx 0.4$ 
solution for various suppression
factors $\omega$ of the Higgs-gluon vertex. The bottom panel shows the 
corresponding Schwinger functions. The lattice data are the same as in fig.\
\ref{fighiggslat}. The solid line is for $\omega=1/4$, the dashed line for
$\omega=1/2$, the dotted line is for $\omega=3/4$ and the dashed-dotted line for
$\omega=1$.}
\label{fighsalt}
\end{figure}

As can be seen from fig.\ \ref{fighiggslat} the result for the Higgs propagator
deviates significantly from the lattice results, the self-energy being
significantly overestimated. While the gluon Schwinger function is reasonably
independent of the truncation, this turns out not to be the case for the Higgs
Schwinger function. In order to show this we study 
the solution for a bare Higgs-gluon vertex suppressed via a scaling factor 
$\omega$.\footnote{In the Landau gauge, due to the transversality of the gluon,
this is equivalent to modifying the tensor structure of the vertex.}
First, we note that the gluon Schwinger function, in contrast to the Higgs one,
is not susceptible to such a change. 

Fig.\ \ref{fighsalt} displays corresponding results for different values of
$\omega$. The position of the first zero tends to increase for decreasing
$\omega$. At $\omega\approx 1/4$  any oscillation, at least within the
available numerical precision, seems to be gone altogether. A fit to the
Schwinger function using (\ref{higgsaltfit}) reveals again additional
structure at very small $t$ which cannot be captured by such a
simple fit. As the fit also does miss some strength at zero momentum for the
propagator this again indicates the presence of further massive contributions
in the propagator.

\subsection{Analytic properties}\label{analytic}

Although the high-temperature limit of the four-dimen\-sional Minkowski theory
is a genuinely Euclidian theory, it will be of interest for other applications
to extract the analytic structure of the propagators investigated.

The gluon propagator exhibits similar behavior for bo\-th solutions but there
are also some significant differences. The denominator of the ansatz
(\ref{gluonfit}) contains two factors, both of which could possibly give rise
to a non-trivial analytic structure. The first part stems from the fit of the
perturbative tail necessary to generate the maximum in the gluon dressing
function. Since all fit parameters are positive, this factor does not give rise to
a pole on the first Riemann sheet. However, it generates a pole on the second
Riemann sheet, which will occur at $1/a_uf$, that is, at Euclidian
momenta. This pole does not have a physical interpretation, and may well be an
artifact of the fit, since (\ref{gluonfit}) is tailored to generate the correct
leading-order perturbative behavior. Thus it generates most likely a structure
which has the Landau pole of perturbation theory on the second Riemann sheet.
We expect that this pole vanishes when using a more sophisticated fit.

The second factor generates a genuine isolated pole at $(-(1+f)/A_z)^{-1/t}$.
This expression has only one value on the first Riemann sheet given by
$(-0.0933+0.1615i)g_3^4$ for the $g=1/2$ and $(-0.1462+0.1248i)g_3^4$ for the
other solution.\footnote{This is of the order of $\Lambda_{QCD}$ when using the 
temperature-scaling of subsection \ref{td}.} In both cases a pole close to the 
origin is generated with an imaginary part larger than the real part in one case. 
For the first solution, the pole is found for an angle significantly above
$\pi/4$ while in the second case somewhat below. In addition, the first
solution generates two more poles on two more Riemann sheets, while the
second solution, with a (most likely) irrational exponent, generates an infinite
number of further poles on an infinite number of Riemann sheets. In both cases,
the residue is complex. In addition, there is a cut along the complete
negative real axis starting at zero. In this way it is similar to the results in 
four dimensions~\cite{Alkofer:2003jj}.

From this analysis we infer that the gluon propagator is positivity violating,
and thus satisfies the requirements of the Kugo-Ojima and Zwanziger-Gribov confinement 
scenario.

\begin{table}
\begin{center}
\begin{tabular}{|c|c|c|}
\hline
Solution & Order & Pole$/g_3^4$ \cr
\hline
$g=1/2$ & Leading & $-0.5112\pm2.191i$ \cr
\hline
$g=1/2$ & Subleading & $-6.11972\pm2.35777i$ \cr
\hline
$g\approx 0.4$ & Leading & $-0.650078\pm2.16822i$ \cr
\hline
$g\approx 0.4$ & Subleading & $-3.69539$ \cr
\hline
\end{tabular}
\end{center}
\caption{Location of the poles of the Higgs propagator for $\omega=1$.}
\label{tabhiggspoles}
\end{table}

On the other hand, the Higgs propagator very likely does not have a branch cut
but a number of simple poles whose locations are given in table
\ref{tabhiggspoles}. The sensitivity to the Higgs-gluon vertex, however,
necessitates further investigations before a firmer conclusion can be drawn.
We only note here that for smaller values of the suppression factor $\omega$
the Higgs propagator agrees better with lattice results, and its poles are then
close to or on the real axis.

\section{Discussion}\label{Interpretation}

Eqs.\ (\ref{fulleqG}), (\ref{fulleqH}) and (\ref{fulleqZ})  describe the
infinite-temperature limit of Landau-gauge Yang-Mills theory. The resulting
gluon propagator exhibits the characteristic behavior of a confined
particle. The Yang-Mills subsector is satisfying both, the Kugo-Ojima 
scenario (\ref{Oehme}) and the Zwanziger-Gribov scenario (\ref{Kugo}), thus describing a confined theory. These properties are very
stable against the assumptions made: the result of confined gluons is nearly
independent of the truncation  and of the properties of the Higgs. Hence, in
accordance with corresponding lattice results, the presence of long-range
chromomagnetic forces is found. 

The chromoelectric sector is simpler in that it is close to a perturbative behavior.
However, as the comparison to lattice calculations shows, at least 
next-to-leading-order perturbative effects or even genuinely
non-perturbative effects play a role. 

Combing these findings with the results from the vacuum~\cite{vonSmekal:1997is}
and from the low-temperature calculations~\cite{torus} has impact on the
understanding of the phase transition. The main difference between the low and
the high-temperature phase is not primarily one between a strongly
interacting and confining system and one with only quasi-free particles.   The
chromoelectric gluons, whose infrared behavior change from over-screening to
screening, comes somewhat close to such a picture. The chromomagnetic gluons
stay over-screened in the infrared and thus confined.  The order parameter for
the phase transition is then necessarily only a chromoelectric one. From the
studies of Wilson loops it is known that only the temporal (electric) Wilson
lines show a behavior typical of deconfinement while the spatial (magnetic)
ones do not~\cite{Bali:1993tz}. Note that the order parameters used to
study the deconfinement transition on the lattice are typically 
chromoelectric ones like the Polyakov lines~\cite{Karsch:2003jg}.
Using the corresponding magnetic Polyakov lines we conjecture that almost
no change will be found.
This point of view is also supported by recent lattice calculations~\cite{Nakamura:2003pu} 
which observe even at $6T_c$ an over-screened magnetic
and a screened electric propagator, albeit on rather small lattices.\footnote{
Based on arguments~\cite{Appelquist:vg} that fermion propagators will vanish in
the infinite temperature limit we do not expect that quarks can change anything
in these considerations.}

The results presented here may have also consequences for the thermodynamic
potential. In the analysis of the Higgs propagator indications  for further
particle-like poles have been found. These could contribute to the pressure.
The way in which over-screened gluons contribute to the energy density but not
to the pressure has been very recently discussed in ref.\ \cite{Zwanziger:2004np}.

Another topic related to the investigations presented here is the recently
established connection between four-dimensional Yang-Mills theory in Coulomb gauge
and th\-ree-dimensional Yang-Mills theory in Landau gauge~\cite{Zwanziger:2003de}.  
The time-time component of the Coulomb-gauge gluon propagator is on the one hand 
directly linked to the static quark-quark potential and on the other hand to the 
connected parts of expectation values of the Faddeev-Popov operator. From this one 
concludes that the potential is approximatively given by the ghost dressing function. 
The infrared behavior of the Coulomb-gauge ghost propagator in four dimensions and 
the Landau-gauge ghost propagator in three dimensions are determined from identical 
equations and one obtains identical infrared exponents. While the solution $g\approx 0.4$
generates a potential which behaves as $\sim 1/k^{3.6}$ and is thus a little less
than linear with distance, the second branch generates a solution proportional
to $1/k^4$ for small momenta, thus generating the behavior expected for a
linear confining potential.

\section{Conclusions and Outlook}\label{Conclusions}

We have analyzed the DSEs of three-di\-mensional Yang-Mills theory with and
without an additional massive adjoint Higgs field and solved them in a given
truncation scheme.  The investigated theory can be regarded as the
high-temperature limit of a four-dimensional Yang-Mills theory. Besides the
propagators we considered the corresponding Schwinger functions. Although the
resulting Higgs function behaves nearly perturbative it does not have a simple
structure and experiences higher-order or even non-perturbative effects. The
chromomagnetic gluons are over-screened. The Faddeev-Popov ghost propagator is
infrared enhanced similar to the four-dimensional one. The corresponding
long-range correlations thus imply confinement of chromomagnetic modes as 
they imply confinement of transverse gluons in four dimensions. 

We have given an approximate expression for the thermodynamic potential. At
the current stage it allows to discriminate which of the two sets of solutions
is the preferred one. Of course, it is our goal to extend our formalism such that
physical observables such as the energy density, pressure and entropy can be
calculated reliably.

The next steps, however, also in connection with the results found in~\cite{torus}, 
will be to include the higher Matsubara frequencies to introduce the effects of a 
finite temperature into the system. This will hopefully allow us to
determine the critical temperature and other aspects of the phase transition. 
Including quarks will allow to address the chiral phase transition. An
extension to finite quark chemical potential is feasible ~\cite{Roberts:2000aa,Epple:dipl},
and thus investigations of the QCD phase diagram based on a calculation of the 
infrared behavior of QCD Green's functions will become possible as well. 

\acknowledgement

The authors are grateful to Christian S. Fischer for his help in 
the early stages of this work. They thank Michael Buballa, Peter Petreczky,
Craig D.\ Roberts, and Daniel Zwanziger for valuable discussions
and Attilio Cucchieri for a critical reading of the manuscript and helpful remarks. 
This work is
supported by the BMBF under grant number 06DA917, by the European Graduate
School Basel-T\"u\-bingen (DFG contract GRK683) and by the
Helmholtz association (Virtual Theory Institute VH-VI-041).

\appendix

\section{Derivation of the Dyson-Schwinger equations}\label{appdse}

The Dyson-Schwinger equations are the equations of motions for a quantum field
theory. They can be derived~\cite{Rivers:hi,Alkofer:2000wg,Roberts:2000aa} 
for the Euclidian version from 
\be
\left(\left(-\frac{\delta S}{\delta \phi^a(x)}|_{\phi=\frac{\delta}{\delta j^a(x)}}
+j^a(x)\right)Z[j^a(x)]\right)_{j=0}=0\label{dsebase}
\ee
where $\phi$ is the field variable, $j$ the corresponding source term and $a$
a generic multi-index. $S$ is the Euclidian action and $Z$ is the generating
functional,
\be
Z=\int{\cal D}\phi^a e^{-S[\phi^a]+\int dx \phi^a(x)j^a(x)}
\ee 
that can be written as $Z=\exp(W)$, $W$ being a function of the external sources. 
The effective action, {\it i.e.\/} the Legendre transform of $W$,  
is defined by
\be
W(j^a)=-\Gamma(\phi^a)+\int d^dxj^a(x)\phi^a(x)
\ee
which implies
\bea
\phi^a=\frac{\delta W}{\delta j^a} \; ,\\
j^a=\frac{\delta\Gamma}{\delta\phi^a} \; .
\eea
In the case of Grassmann fields, $c$, such as ghosts and fermions, two independent
sources are necessary. This modifies the above to
\bea
&Z=\int{\cal D}c^a{\cal D}\bar c^a e^{-S[c^a,\bar c^a]+
\int d^dx(\bar\eta^a(x) c^a(x)+\bar c^a(x)\eta^a(x))}\\
&c^a(x)=\frac{\delta W}{\delta\bar\eta^a(x)}
\quad\quad\quad\quad\bar c^a(x)=-\frac{\delta W}{\delta\eta^a(x)}\nonumber\\
&W(\eta^a,\bar\eta^a)\nonumber\\
&=-\Gamma(u^a,\bar u^a)+\int d^dx(\bar\eta^a(x)c^a(x)+\bar c^a(x)\eta^a(x))\\
&\eta^a(x)=\frac{\delta\Gamma}{\delta\bar c^a(x)}\quad\quad\quad\quad
\bar\eta^a(x)=-\frac{\delta\Gamma}{\delta c^a(x)}\nonumber
\eea
where all derivatives with respect to Grassmann variables act in the direction
of ordinary derivatives.

The general procedure to obtain the corresponding Dyson-Schwinger equations is
to calculate the expression (\ref{dsebase}) for a given action and then 
perform once more a functional derivative with respect to the field or with
respect to the conjugate field in case of anti-commuting fields. The additional
source term then yields the propagator while the right-hand-side of the
equations are found by the derivative of the action.

In general, propagators and their inverse are defined as
\bea
&\frac{\delta^2\Gamma}{\delta\phi^b(y)\delta\phi^a(x)}&=D^{ab}(x-y)^{-1}
\label{geninvprop}\\
&\frac{\delta^2 W}{\delta\phi^b(y)\delta\phi^a(x)}&=D^{ab}(x-y)\label{genprop}
\eea
while full $n$-point vertices are defined by a $n$-fold derivative of $\Gamma$.
For the ghost-gluon vertex which is defined as
\be
\frac{\delta^3\Gamma}{\delta c^a(x)\delta\bar c^b(y)\delta A^c_\mu(z)}
=\Gamma_\mu^{c\bar cA;abc}(x,y,z) \; ,
\ee
the sequence of derivatives is relevant (in contrast to the gluon vertex
functions). 

\begin{figure}
\epsfig{file=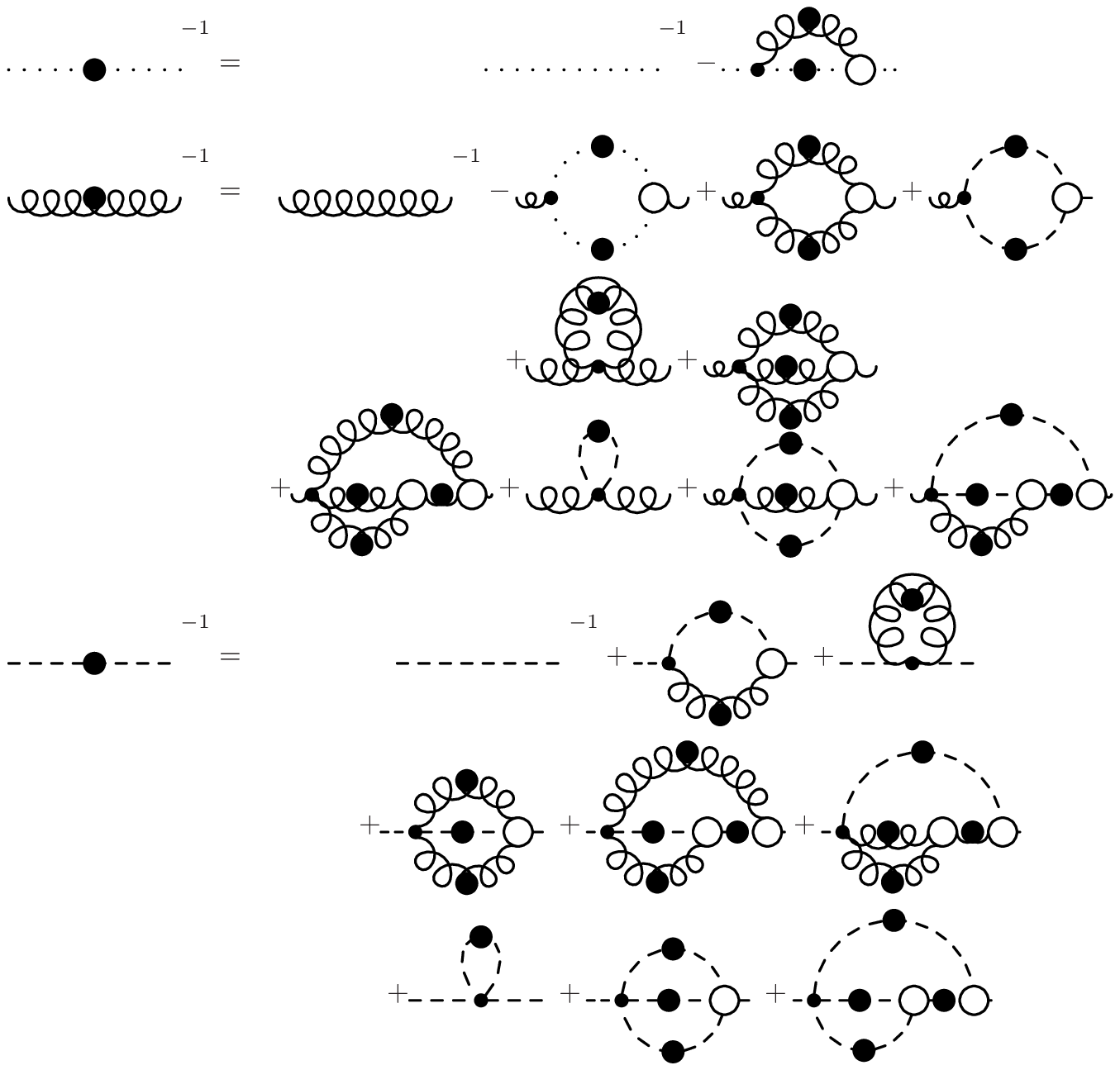,width=0.95\linewidth}
\caption{The DSEs of the propagators of the Yang-Mills theory with an adjoint
Higgs given in (\ref{tlgheq}), (\ref{tlgleq}), and (\ref{tlhiggseq}). The same
conventions apply as in Fig.\ \ref{threedsys}.}
\label{figymhfull}
\end{figure}

Using the identities
\bea
&\frac{\delta^2 W}{\delta j^e_\mu(x)\delta\bar\eta^d(x)}\nonumber\\
&=-\int d^dzd^dw\frac{\delta^2 W}{\delta j_\nu^f(z)\delta j^e_\mu(x)}
\frac{\delta^2\Gamma}{\delta\bar c^g(w)\delta A_\nu^f(z)}\frac{\delta^2 W}
{\delta\eta^g(w)\delta\bar\eta^d(x)}\label{dseid1}\\
&\pdm^x\frac{\delta}{\delta\eta^c(x)}=\int d^dz \Big(\pdm^x\frac{\delta\Gamma}
{\delta\eta^c(x)\delta\bar c^e(z)}\Big)\frac{\delta}{\delta\eta^e(z)}
\label{dseid2}\\
&\frac{\delta^3 W}{\delta A_\nu^b(y)\delta j^c_\sigma(x)\delta^e_\sigma(x)}
=\nonumber\\
&-\int dzdw \frac{\delta^2 W}{\delta j_\sigma^c(x)\delta j_\rho^f(z)}
\frac{\delta^3\Gamma}{\delta A_\nu^b(y)\delta A_\rho^f(z)\delta A_\omega^g(w)}
\frac{\delta^2 W}{\delta j_\omega^g(w)\delta j_\sigma^e(x)},\label{dseid3}
\eea
it is straightforward (although tedious) to derive the DSEs in position space. 
Performing a Fourier transformation to momentum space, where all momenta are
defined as incoming and momentum conservation at the vertices is taken into
account, the DSEs in momentum space are derived. To make the appearance of the
tree-level vertices explicit, it is necessary to add in general permuted
versions of each diagram and using the (anti-)symmetry of the vertices in their
respective fields. The ghost propagator equation reads
\bea
&D^{ab-1}_G(p)=-\delta^{ab}p^2\nonumber\\
&+\int\frac{d^dq}{(2\pi)^d}\Gamma_\mu^{tl;c\bar cA;dae}(-q,p,q-p)
D^{ef}_{\mu\nu}(p-q)D^{dg}_G(q)\cdot\nonumber\\
&\cdot\Gamma^{c\bar cA;bgf}_\nu(-p,q,p-q) \; ,\label{tlgheq}
\eea
the gluon one is given by
\bea
&D^{ab-1}_{\mu\nu}(p)=\delta^{ab}(\delta_{\mu\nu}p^2-p_\mu p_\nu)\nonumber\\
&-\int\frac{d^dq}{(2\pi)^d}\Gamma_\mu^{tl;c\bar cA;dca}(-p-q,q,p)
 D_G^{cf}(q) D_G^{de}(p+q)\cdot\nonumber\\
&\cdot\Gamma_\nu^{c\bar c A;feb}(-q,p+q,-p)\nonumber\\
&+\frac{1}{2}\int\frac{d^dq}{(2\pi)^d}
\Gamma^{tl;A^3;acd}_{\mu\sigma\chi}(p,q-p,-q)
D^{cf}_{\sigma\omega}(q)D^{de}_{\chi\lambda}(p-q)\cdot\nonumber\\
&\cdot\Gamma^{A^3;bfe}_{\nu\omega\lambda}(-p,q,p-q)\nonumber\\
&+\frac{1}{2}\int\frac{d^dq}{(2\pi)^d}
\Gamma_\mu^{tl;A\phi^2;acd}(p,q-p,-q)D^{de}(q)D^{cf}(p-q)\cdot\nonumber\\
&\cdot\Gamma^{A\phi^2;bef}_\nu(-p,q,p-q)\nonumber\\
&+\frac{1}{2}\int\frac{d^d q}{(2\pi)^d}
\Gamma_{\mu\nu\sigma\rho}^{tl;A^4;abcd}(p,-p,q,-q)D^{cd}_{\sigma\rho}(q)
\nonumber\\
&+\frac{1}{6}\int\frac{d^dqd^dk}{(2\pi)^{2d}}
\Gamma_{\mu\sigma\xi\chi}^{tl;A^4;acde}(p,-q,-p+q-k,k)\cdot\nonumber\\
&\cdot D_{\chi\lambda}^{dh}(p-q-k)D_{\xi\rho}^{cf}(q)D_{\sigma\omega}^{eg}(k)
\Gamma^{A^4;hbfg}_{\lambda\nu\rho\omega}(p-q-k,-p,q,k)\nonumber\\
&+\frac{1}{2}\int\frac{d^dqd^dk}{(2\pi)^{2d}}
\Gamma^{tl;A^4;acde}_{\mu\delta\gamma\sigma}(p,-q,q-k-p,k)
D_{\gamma\lambda}^{dh}(k+p-q)\cdot\nonumber\\
&\cdot\Gamma_{A^3;\nu\rho\omega}^{bfg}(-p,p+k,-k)
\Gamma_{A^3;\lambda\chi\xi}^{hij}(k+p-q,q,-k-p)\cdot\nonumber\\
&\cdot D_{\rho\xi}^{fj}(p+k)D_{\sigma\omega}^{eg}(k)D_{\delta\chi}^{ci}(q)
\nonumber\\
&+\frac{1}{2}\int\frac{d^dq}{(2\pi)^d}
\Gamma^{tl;A^2\phi^2;abdf}_{\mu\nu}(p,-p,q,-q)D^{df}(q)\nonumber\\
&-\frac{1}{2}\int\frac{d^dqd^dk}{(2\pi)^{2d}}
\Gamma_{\mu\sigma}^{tl;A^2\phi^2;aedf}(p,q-p-k,-q,k)D^{fi}(k)\nonumber\\
&\cdot D_{\sigma\rho}^{eg}(p-q+k)D^{dh}(q)
\Gamma_{\rho\nu}^{A^2\phi^2;gbhi}(p-q+k,-p,q,-k)\nonumber\\
&+\int\frac{d^dqd^dk}{(2\pi)^{2d}}
\Gamma_{\mu\sigma}^{tl;A^2\phi^2;aedf}(p,q-p-k,-q,k)D_{\mu\rho}^{eg}(p+k-q)
\cdot\nonumber\\
&\cdot D^{fi}(k)D^{hk}(k+p)D^{dj}(q)
\Gamma^{A\phi^2;gjk}_\rho(p+k-q,q,-k-p)\cdot\nonumber\\
&\cdot\Gamma^{A\phi^2;bhi}_\nu(-p,k+p,-k) \; ,\label{tlgleq}
\eea
and finally the one for the Higgs is  
\bea
&D^{ab}(p)^{-1}=\delta^{ab}(p^2+m_h^2)\nonumber\\
&+\int\frac{d^dq}{(2\pi)^d}\Gamma^{tl;A\phi^2;eac}_\nu(-p-q,p,q)
D_{\nu\mu}^{cg}(p+q)\cdot\nonumber\\
&\cdot D^{fc}(q)\Gamma_\mu^{gbf}(p+q,-p,-q)\nonumber\\
&+\frac{1}{2}\int\frac{d^dq}{(2\pi)^d}
\Gamma_{\mu\nu}^{tl;A^2\phi^2;cdab}(q,-q,p,-p)D_{\mu\nu}^{cd}(q)\nonumber\\
&-\frac{1}{2}\int\frac{d^dqd^dk}{(2\pi)^d}
\Gamma_{\mu\sigma}^{tl;A^2\phi^2;cdae}(-p-q+k,-k,p,q)\cdot\nonumber\\
&\cdot D_{\mu\nu}^{cg}(p+q-k)D_{\rho\sigma}^{id}(k)D^{eh}(q)\cdot\nonumber\\
&\cdot\Gamma_{\rho\nu}^{A^2\phi^2;igbh}(-k,p+q-k,-p,q)\nonumber\\
&+\frac{1}{2}\int\frac{d^dqd^dk}{(2\pi)^{2d}}
\Gamma_{\mu\sigma}^{tl;A^2\phi^2;cdae}(-q,k,p,-p+q-k)D_{\mu\nu}^{cg}\cdot
\nonumber\\
&\cdot D^{ej}(p-q+k)(q)D_{\rho\sigma}^{id}(k)
\Gamma_\nu^{A\phi^2;gjk}(q,p-q+k,-p-k)\cdot\nonumber\\
&\cdot D^{kh}(p+k)
\Gamma_\rho^{A\phi^2;ibh}(-k,-p,p+k)\nonumber\\
&+\frac{1}{2}\int\frac{d^dqd^dk}{(2\pi)^{2d}}
\Gamma_{\mu\chi}^{tl;A^2\phi^2;cdae}(-q,q+k,p,-p-k)D^{cg}_{\mu\nu}(q)\cdot
\nonumber\\
&\cdot D^{eh}(p+k)D_{\lambda\chi}^{kd}(q+k)
\Gamma_{\sigma\nu\lambda}^{A^3;jgk}(k,q,-q-k)\cdot\nonumber\\
&\cdot D_{\rho\sigma}^{ij}(k)\Gamma_\rho^{A\phi^2;ibh}(-k,-p,p+k)\nonumber\\
&+\frac{1}{2}\int\frac{d^dq}{(2\pi)^d}\Gamma^{tl;\phi^4;abcd}(p,-p,q,-q)
D^{cd}(q)\nonumber\\
&-\frac{1}{6}\int\frac{d^dqd^dk}{(2\pi)^{2d}}
\Gamma^{tl;\phi^4;agch}(p,-p+q-k;k;-q)D^{he}(q)\cdot\nonumber\\
&\cdot D^{gd}(p-q+k)D^{fc}(k)\Gamma^{\phi^4;debf}(p-q+k,q,-p,-k)\nonumber\\
&+\frac{1}{3}\int\frac{d^dqd^dk}{(2\pi)^{2d}}
\Gamma^{tl;\phi^4;aicj}(p,-p-k+q,k,-q)D^{id}(p+k-q)\cdot\nonumber\\
&\cdot D^{jg}(q)D^{fc}(k)\Gamma^{\phi^3;gdh}(q,p+k-q,-k-p)\cdot\nonumber\\
&\cdot D^{eh}(p+k)\Gamma^{\phi^3;ebf}(p+k,-p,-k) \; .\label{tlhiggseq}
\eea
The tree-level vertices for the ghost-gluon-, 3-gluon-, 4-gluon-, 
2-gluon-Higgs-, 2-gluon-2-Higgs- and 4-Higgs- interactions ha\-ve been used.
They are given by
\bea
\Gamma_\mu^{tl;c\bar cA;abc}(p,q,k)&=&ig_3f^{abc}q_\mu\label{tlcca}\\
\Gamma_{\mu\nu\rho}^{tl;A^3;abc}(p,q,k)&=&-ig_3f^{abc}((q-k)_\mu\delta_{\nu\rho}
+(k-p)_\nu\delta_{\mu\rho}\nonumber\\
&&+(p-q)_\rho\delta_{\mu\nu})\label{tlggg}\\
\Gamma_{\mu\nu\sigma\rho}^{tl;A^4;abcd}(p,q,k,l)&=&g_3^2(f^{eab}f^{ecd}
(\delta_{\mu\sigma}\delta_{\nu\rho}-\delta_{\mu\rho}\delta{\nu\sigma})
\nonumber\\
&&+f^{gac}f^{gbd}(\delta_{\mu\nu}\delta_{\sigma\rho}-\delta_{\mu\rho}
\delta_{\nu\sigma})\nonumber\\
&&+f^{gad}f^{gbc}(\delta_{\mu\nu}\delta_{\sigma\rho}-\delta_{\mu\sigma}
\delta_{\nu\rho}))\label{tl4g}\\
\Gamma_\mu^{tl;A\phi^2;abc}(p,q,k)&=&ig_3f^{abc}(q-k)_\mu\label{tlgh}\\
\Gamma_{\mu\nu}^{tl;A^2\phi^2;abcd}(p,q,k,l)&=&g_3^2
\delta_{\mu\nu}(f^{eac}f^{ebd}+f^{ead}f^{ebc})\label{tl2g2h}\\
\Gamma^{tl;\phi^4;abcd}(p,q,k,l)&=&2h(\delta_{ab}\delta_{cd}+\delta_{ac}
\delta_{bd}+\delta_{ad}\delta_{bc})\label{tl4h},
\eea
where the momentum conserving $\delta$-functions have been suppressed. 
The complete graphical representation of (\ref{tlgheq}), (\ref{tlgleq}), and 
(\ref{tlhiggseq}) is shown in Fig.\ \ref{figymhfull}.

\section{Integral kernels}\label{appkernels}

The integral kernels in eqs. (\ref{fulleqG}),  (\ref{fulleqH}) and  
(\ref{fulleqZ}) are obtained by using tree-level vertices and 
performing the corresponding contractions. The ghost kernel is given by
\be
A_t(k,q,\theta)=-\frac{q^2\sin^3(\theta)}{(k^2+q^2-2kq\cos\theta)^2}.
\ee
The contributions in the Higgs equation are
\bea
N_1(k,q,\theta)=-\frac{2q^2\sin^3(\theta)}{(k^2+q^2+2kq\cos\theta)^2}\\
N_2(k,q,\theta)=-\frac{2\sin^3(\theta)}{k^2+q^2+2kq\cos\theta}.
\eea
The kernels in the gluon equations are finally
\bea
&R(k,q,\theta)=-\frac{((\zeta-1)kq\cos(\theta)-q^2+\zeta q^2\cos^2(\theta))
\sin\theta}{2k^2(k^2+q^2+2kq\cos\theta)}\\
&M_L(k,q,\theta)=\nonumber\\
&\frac{((\zeta-1)(k^2+4kq\cos\theta)-4q^2+4q^2\zeta\cos^2(\theta))\sin\theta}
{4k^2(k^2+q^2+2kq\cos\theta)}\\
&M_T(k,q,\theta)=\frac{\sin\theta}{4k^2(k^2+q^2+2kq\cos\theta)}\cdot\nonumber\\
&\cdot\Big((k^2+2q^2)((\zeta-9)k^2-4q^2)+8(\zeta-3)(k^2+q^2)kq\cos\theta
\nonumber\\
&+(8\zeta q^4+(\zeta+7)k^4+4(5\zeta-1)k^2q^2)\cos^2(\theta)\nonumber\\
&+4(4\zeta q^2+(\zeta+3)k^2)\cos^3(\theta)+4\zeta k^2q^2\cos^4(\theta)\Big)
\eea
The modified gluon vertex is introduced by multiplying  $M_T$ with (\ref{g3vertex}).

\section{Perturbative expressions}\label{appUV}

Replacing all full quantities in the truncated DSEs 
(\ref{fulleqG}),  (\ref{fulleqH}) and  (\ref{fulleqZ}) 
by their tree-level values, {\it i.e.\/} one for all dressing functions and 
the tree-level vertices (\ref{tlcca}-\ref{tl4h}) instead of the full vertices, 
the standard perturbation theory to one-loop order is obtained. 
It is then possible to calculate the leading-order perturbative dressing 
functions. For the ghost this becomes
\be
G(k)^{-1}=1-\frac{g_3^2C_A}{16k}\label{pertgh}
\ee
to leading order, independent on the presence of a Higgs. Note that a 
Landau pole at $k={\cal O}(g_3^2)$ is present.

The calculation of the gluon self-energy is a little more tricky since, although
finite in three dimensions due to a Slav\-nov-Taylor identity, each single
contribution is linearly divergent. 
Corresponding problems are most easily circumvented when
contracting the gluon equation with the Brown-Penning\-ton projector.
Performing the calculation in pure Yang-Mills theory yields
\be
Z(k)^{-1}=1-\frac{11g_3^2C_A}{64k}\label{pertgl}.
\ee
Including the Higgs one obtains
\bea
&Z(k)^{-1}=1-\frac{11g_3^2C_A}{64k}\nonumber\\
&+\frac{g_3^2C_A}{16\pi k}\Big(2\frac{m_h}{k}-\frac{k^2+4m_h^2}{k^2}\csc^{-1}
\sqrt{1+\frac{4m_h^2}{k^2}}\Big).\label{pertglwh}
\eea
Since non-perturbative effects already arise at order $1/k^2$ in the present
truncation scheme, the only interesting part is for $k\gg g_3^2,m_h$ 
leading to
\be
Z(k)^{-1}=1-\frac{9g_3^2C_A}{64k}\label{pertgllead}.
\ee
Finally the Higgs self-energy is
\bea
&H(k)^{-1}=1+\frac{m_h^2}{k^2}+\frac{g_3^2C_A}{k}\frac{m_h}{4\pi k}\nonumber\\
&+\frac{g_3^2C_A}{k}\frac{1}{2\pi k} \left(-m_h+\left(\frac{m_h^2+k^2}{k}-2k\right)\arcsin
(\sqrt{\frac{k^2}{m_h^2+k^2}})\right)\label{perth}
\eea
\noindent where the third term is the tadpole contribution. The leading contribution is
\be
H(k)^{-1}=1-\frac{g_3^2C_A}{4k}.\label{perthlead}
\ee
The coupling constant $h$ is determined by requiring that the sum of
the tadpole kernels already generate a finite integral, {\it i.e.} independent of the
regularization scheme. This is obtained by requiring (\ref{hvalue}).

Note that, in three dimensions, resummation has only effects from order $g_3^4$
on since, by dimensional arguments alone, only tree-level expressions in the
loops can contribute at order $g_3^2$. Therefore (\ref{pertgh}), 
(\ref{pertgllead}) and (\ref{perthlead}) already constitute the resummed
solution. This is confirmed by the numerical calculations presented in section
\ref{Numerical}.

Note further that this also ensures that gauge symmetry is intact to one-loop
order in the regime of applicability of leading-order resummed perturbation
theory. The violation of gauge symmetry in this approach is, at least, not
stronger than in ordinary leading-order perturbation theory.

\section{Tadpoles}\label{appTadpoles}

The tadpoles used for the ghost-loop-only system read:
\bea
T^{GG}=&-\frac{g_3^2C_A}{(2\pi)^2}\int dqd\theta\Big(R_D(k,q)(G(q)G(k+q)\nonumber\\
&-A_g^2q^{-2g}(k+q)^{-2g})+R_3(k,q)\big(G(q)G(k+q)\nonumber\\
&-A_g^2q^{-2g}(k+q)^{-2g}-1\big)\Big)\label{ggtad1}.
\eea
\noindent Here the kernel $R$ has been split into its convergent, 
$\zeta$-in\-dependent part $R_0$, its finite $\zeta$-dependent part $R_3$ and its 
divergent part $R_D$ as
\be
R=R_0+R_3+R_D.\label{rsplit}
\ee
\noindent For the pure Yang-Mills theory this is altered to
\bea
T^{GG}=&-\frac{g_3^2C_A}{(2\pi)^2}\int dqd\theta\Big(R_D(k,q)(G(q)G(k+q)\nonumber\\
&-A_g^2q^{-2g}(k+q)^{-2g})\nonumber\\
&+R_3(k,q)\big(G(q)G(k+q)-A_g^2q^{-2g}(k+q)^{-2g}\nonumber\\
&-1\big)+M_{TD}(k,q)Z(q)Z(k+q)\Big)\label{ggtad2},
\eea
\noindent where $M_{TD}$ contains the divergent part of $M_T$, including the 
alterations due to the dressed 3-gluon-vertex (\ref{g3vertex}). For the full theory 
the additional tadpole in the gluon equation is given by
\be
T^{GH}=-\frac{g^2C_A}{(2\pi)^2}\int dqd\theta\Big(M_{LD}(k,q)H(q)H(k+q)\Big).\label{ghtad}
\ee
\noindent In the Higgs equation, the tadpoles are set to
\be
T^{HG}+T^{HH}=\frac{g_3^2C_A}{k^2}\frac{m}{4\pi}=:\frac{\delta m^2}{k^2}.\label{higgstadpole}
\ee
\noindent A detailed account for the construction of the tadpoles is given in 
ref.\ \cite{Maas:phd}.

\section{Infrared expressions}\label{appIR}

As already argued in section \ref{Infrared}, the only solution without further
assumption is that of ghost dominance. This then requires the calculation
of $I_{GT}$ and $I_{GG}$ in (\ref{ghostir}) and (\ref{gluonir}) only. The latter 
can be obtained straightforwardly when using the ansatz (\ref{iransatz}) and the
general formula
\bea
&\int\frac{d^dq}{(2\pi)^d}x^\alpha z^\beta\nonumber\\
&=\frac{1}{(4\pi)^{\frac{d}{2}}}\frac{\Gamma(-\alpha-\beta-\frac{d}{2})
\Gamma(\frac{d}{2}+\alpha)\Gamma(\frac{d}{2}+\beta)}{\Gamma(d+\alpha+\beta)
\Gamma(-\alpha)\Gamma(-\beta)}y^{\frac{d}{2}+\alpha+\beta}\label{dimregrule}
\eea
valid for finite integrals, where $x=q^2$, $z=(q-k)^2$ and $y=k^2$. It yields
\bea
&I_{GG}=-\frac{g_3^2C_A\pi}{(4\pi)^{\frac{d}{2}}}\cdot\nonumber\\
&\cdot\frac{2^{4g-2d}(d-4g)(2+d(\zeta-2)-4g(\zeta-1)-\zeta)\Gamma(d-2g)
\Gamma(2g-\frac{d}{2})}{(d-1)g^2\Gamma(\frac{1+d-2g}{2})^2\Gamma(g)^2}
\eea
Note that the integral is convergent iff
\be
\frac{d-1}{2}\ge g>\frac{d-2}{4}\label{dglimits}
\ee
where the equality on the upper boundary requires the result to exist only in
the sense of a distribution.

Performing the same calculation for the ghost self-energy is more complicated.
Since it is, in general, a  divergent quantity Eq.\ (\ref{dimregrule}) cannot
be applied directly. It is necessary to regularize and then renormalize the
expression. This can be done in a momentum subtraction scheme~\cite{Zwanziger:2001kw} 
or via dimensional regularization. Here, the later has been performed by applying 
the standard rules of dimensional regularization~\cite{Peskin:ev}. This then 
immediately gives a finite result. However, by doing so a
divergent quantity has been removed which is formally eliminated by setting
$-\widetilde Z_3$ equal to this quantity. This procedure yields the same
result as the momentum subtraction scheme. The range allowed for $g$ in
(\ref{dglimits}) renders $I_{GT}$ not only to have a divergence of logarithmic
or linear order in even or uneven dimensions, but also quadratic or cubic
divergences. Using a subtraction scheme, it would be necessary to include the
next term in the Taylor expansion. On the other hand, dimensional
renormalization directly yields
\bea
&I_{GT}=\frac{g_3^2C_A}{(4\pi)^{\frac{d}{2}}}\cdot\nonumber\\
&\cdot\frac{2^{1-2g}(4^g(d-3)d+2^{1+2g}(1+g-dg))\Gamma(\frac{d}{2}-g)
\Gamma(-g)\Gamma(2g)}{(2-d+2g)(d+2g)\Gamma(\frac{d}{2}-2g)\Gamma(g)
\Gamma(\frac{d}{2}+g)}
\eea
Note that this expression becomes negative already for values allowed by
(\ref{dglimits}), {\it e.g.\/} for $g\ge3/4$ in three dimensions, thus 
reducing the allowed range and leading to the plots in Fig.\ \ref{figexp}.

\end{document}